\documentstyle[11pt,aaspp4,flushrt,astrobib]{article}  

\newcommand{\be}{\begin{equation}}
\newcommand{\ee}{\end{equation}}
\newcommand{\bea}{\begin{eqnarray}}
\newcommand{\eea}{\end{eqnarray}}
\newcommand{\eq}[1]{equation~(\ref{eq-#1})}

\newcommand{\fig}[1]{figure~\ref{fig-#1}}
\newcommand{\Fig}[1]{Figure~\ref{fig-#1}}

\newcommand{\area}{{\mathcal S}}
\newcommand{\qfloor}{{\lfloor q/2 \rfloor}}
\newcommand{\ri}{r_i}
\newcommand{\ro}{r_o}
\newcommand{\total}{{\mbox{\scriptsize total}}}
\newcommand{\src}{{\mbox{\scriptsize src}}}
\newcommand{\back}{{\mbox{\scriptsize back}}}
\newcommand{\inn}{{\mbox{\scriptsize inner}}}
\newcommand{\out}{{\mbox{\scriptsize outer}}}
\newcommand{\cen}{{\mbox{\scriptsize src}}}
\newcommand{\rinn}{r_\inn}
\newcommand{\rout}{r_\out}
\newcommand{\rcen}{r_\cen}
\newcommand{\binn}{B_\inn}
\newcommand{\bout}{B_\out}
\newcommand{\bcen}{B_\cen}

\slugcomment{{\tt LA-UR 98-3076}}
\lefthead{Theiler and Bloch}
\righthead{Multiple concentric annuli\ldots}

\begin{document}

\title{%
Multiple concentric annuli for characterizing\\
spatially nonuniform backgrounds
}
\author{James Theiler and Jeff Bloch}
\affil{Space and Remote Sensing Sciences Group, NIS-2, MS-D436,\\
	Los Alamos National Laboratory, Los Alamos, NM 87545}

\renewcommand{\baselinestretch}{1}\large\normalsize

\begin{abstract}
A method is presented for estimating the background at a given location
on a sky map by interpolating the estimated background from a set of
concentric annuli which surround this location.  If the background
is nonuniform but smoothly varying, this method provides a more accurate
(though less precise) estimate than can be obtained with a 
single annulus.  Several applications of multi-annulus background 
estimation are discussed, including direct testing for
point sources in the presence of a nonuniform background,
the generation of ``surrogate maps''
for characterizing false alarm rates, and precise testing of the null
hypothesis that the background is uniform.
\end{abstract}


\section{Introduction}

This study is motivated by the search for rare bright transient point
sources in count-limited sky maps generated from data taken by the
ALEXIS (Array of Low Energy X-ray Imaging Sensors) satellite.  
Each of the six
telescopes in the array uses a spherical narrow-band normal-incidence 
multi-layer mirror with a micro-channel plate detector at the prime
focus to detect individual EUV photons over a wide
33$^o$ field \cite{A_Pried_SPIE_88a,A_Bloch_Rollercoaster_93a}.  The data are
taken in a scanning survey mode, with almost half of the sky (the
anti-solar hemisphere) covered every fifty seconds.  Individual
photons are time-tagged, and after accounting for spacecraft attitude
as a function of time \cite{1997JGCD...20.1033P}, each photon is
identified with a position on the sky.  In this original form, the
``map'' is a photon event list, but in practice, these photons are
binned into square pixels in a Hammer-Aitoff or a
quadrilateralized spherical cube projection~\cite{Greisen95,Greisen96} of the
sky.  For a given candidate point source location, we estimate a
source strength by counting photons in a small region of the map about
the size of the telescope point spread function (more accurate
estimates can be obtained by convolving the photon counts with the
point spread function), and comparing that to an estimate of the
background.

If quantitative models are available for all of the various noise
sources in the detector and on the sky, then {\em ab initio} estimates
of the background can be made.  To be useful, these models must be
accurate and reliable, and this is not always possible; for ALEXIS,
one of the main reasons for going into space was to
identify and quantify these various backgrounds for this new multi-layer
instrument technology.  A simple
alternative is to estimate the background
by looking at count rates in the vicinity of the candidate point
source
({\it e.g.,} see \citeNP{EUVE2Catalog}).
Annular regions are particularly useful, since they are
insensitive to linear gradients in the background, but there is 
still an implicit assumption that the background is
spatially uniform (or at worst has only linear variation) 
over the area that includes the point
source location and the background estimation region.

With the ALEXIS data, the spatial nonuniformity of the
background is quite evident (see \fig{image}a), and arises
from a number of effects: among these are uneven exposure of the
scanning telescopes, telescope vignetting, 
pinhole leaks, 
detector masks, 
time-varying high energy penetrating particle flux,
an anomalous background component
which varies with the angle between the look direction
and the spacecraft velocity vector
\cite{1994SPIE.2280..297B}, 
motion of the moon across the field, and possibly even the spatial
structure inherent in the diffuse X-ray sky.
In this study, we are not concerned with
nonuniformities arising from source confusion since point sources
are relatively sparse at the ALEXIS sensitivity.
Most of these effects
are approximately known, and work is in progress to more accurately
model them.  However, our approach for point source detection is to
estimate background from the count map itself.

In this article, we will describe the use of multiple concentric
annuli for characterizing a spatially nonuniform background.  The most
direct application is the estimation of background at a candidate
point source location.  The multiple-annulus approach is
mathematically equivalent to fitting the background with a
two-dimensional polynomial of order $q=2n-1$, where $n$ is the number
of annuli.  Direct fits, even for relatively low order $q$, can be
unwieldy, but by integrating these polynomials over concentric annuli, we
find that the interpolation procedure can be considerably simplified.
With two annuli, for example, we effectively fit a cubic polynomial, but
we compute only two coefficients instead of ten.

If multiple-annulus estimates of
the background can be made more accurate, then a point source detection
algorithm that is based on these better estimates will be more 
sensitive to real point
sources for a fixed rate of false alarms.  A low false alarm rate is 
always desirable, but what is particularly
important is that the false alarm rate be well calibrated.
We show how this can be done using multiple-annulus methods to estimate a
smooth but nonuniform estimate of the background.  From this
background, one then generates a Monte-Carlo 
ensemble of ``surrogate'' maps, and by applying the point source detection
algorithm to these maps (which have no point sources), one can estimate
the false alarm rate.  Finally, we will show how 
multiple annuli can also be
used to characterize the magnitude of the background nonuniformity.  In
particular, we will describe a test of the null hypothesis that the
background is uniform.  The test, based on counts in concentric
annuli, is especially sensitive to the nonuniformities that lead to
poor estimates of the background level at the center of the annuli,
yet is completely insensitive to linear gradients which have no effect
on background estimates at the center of the annuli.

In Section 2, we will introduce notation, and then derive the linear
combination of background annuli counts that provides an unbiased estimator
for background in a source region.  We will consider in particular the 
use of two annuli, the simplest and probably most useful case, as well
as the limit of an infinite number of annuli.
In Section~3, we will derive the
bias and variance (a.k.a. accuracy and precision) 
of multiple annulus estimators, compare
the performance of single and multiple annulus methods, derive optimal
annulus partitions, and suggest heuristic algorithms which trade
off bias and variance error.  In Section~4, we will illustrate the use
of multiple annuli on two example data sets, one real and one artificial.
In that section, we will provide four different applications of multiple
annuli estimators:  estimating background level,
detecting point sources, generating surrogate maps, and quantifying
background nonuniformity.

\section{Background estimation with multiple annuli}
\label{sec:backest}

In this section, we will derive expressions for the average background in
a source region as a linear function of the backgrounds integrated
over annular regions surrounding the source.  We will derive
separate expressions for square and circular regions, but both
will have the property that the coefficients of the linear
combination depend only on the geometry of the annuli.

For a given region $\area$ of a sky map, we will use $N_\area$ to denote
the observed number of counts (photons) in the region, and ${\cal N}_\area$ to
denote the {\em expected} number of counts due only to the smoothly
varying nonuniform background.  These are dimensionless ``counts'' and
should not be confused with the count ``rate'' (photons per unit time).
For a given position
$(x,y)$ on the sky map, let $B(x,y)$ denote the ``count density''
of the background -- expected counts per unit area at that position.  Then,
\be
	{\cal N}_\area = \int\!\!\!\int_\area B(x,y) \,dx\,dy.
	\label{eq-mu}
\ee
The average count density in a region $\area$ is called
$B_\area$ and is given by 
\be
	B_\area = 
	\frac{{\cal N}_\area}{A_\area} =
	\frac{\int\!\!\int_\area B(x,y)\,dx\,dy}{\int\!\!\int_\area dx\,dy}
	\label{eq-B-define}
\ee
where $A_\area$ is the area of region $\area$.  

Here, ${\cal N}_\area$ is the ``true'' background level.  If there is
only background (no point sources) in $\area$, then the actual number
of observed counts $N_\area$ will be Poisson-distributed with mean
${\cal N}_\area$.  For example, if $\area$ is the annular region
around the candidate point source, then we will usually estimate the
background level with $\hat B_\area = \hat{\cal N}_\area/A_\area = 
N_\area/A_\area$.

In what follows, we will assume that the background count density
$B(x,y)$ can be accurately approximated,
over a small region around the potential point source,
by an order-$q$ Taylor series polynomial\footnote{We recognize that
even large order polynomials will not model abrupt discontinuities
in the background -- these, alas, are all too common in actual practice.
Potential point sources near background discontinuities must be treated
with special care.}:
\be
	B(x,y) = \sum_{n+m \le q} c_{nm}x^ny^m.
	\label{eq-B-taylor}
\ee

The formalism that we will develop applies to both square and circular
annuli.  Square annuli are more conveniently employed in maps built from
square pixels, while circular annuli are more convenient if
the data are recorded in a photon event list.
We will characterize these annuli in terms of moments, $R_k$, defined
as follows:  
Let $\ri$ and $\ro$ denote
the inner and outer radii of the annulus (a disk is just an annulus with
$\ri=0$).  It will turn out that only
the even moments are important, so we will {\em define}
\bea
	R_k(\ri,\ro) &\equiv& \frac{1}{k+1}\,
	\frac{\ro^{2k+2} - \ri^{2k+2}}{\ro^2 - \ri^2} 
	\label{eq-define-Rk} \\
	&=& \frac{1}{k+1}\sum_{j=0}^k \ro^{2j}\ri^{2k-2j}.
\eea
For circular annuli, this corresponds precisely to the $2k$-th moment 
$\langle r^{2k}\rangle$.  For square annuli, there is a $k$-dependent
prefactor which is not important for our purposes.

\subsection{Average counts in a square annulus}

Let $\area$ be a square of side $2r$ ({\it i.e.,} of ``radius'' $r$).
Then, from \eq{mu} and \eq{B-taylor}, we can write the expected number of
counts in the square as
\be
	{\cal N}_\area = \sum_{n+m \le q} c_{nm} \int_{-r}^r x^n\,dx \int_{-r}^r y^m\,dy.
	\label{eq-C-square}
\ee
Since the square is symmetric in both $x$ and $y$, it follows that
$\int_{-r}^r dx\,x^n\,\int_{-r}^r dy\, y^m$ 
is zero unless both $n$ and $m$ are even.  So
we need only perform the sum over even indices; that is:
\bea
{\cal N}_\area &=& \sum_{2n+2m \le q} c_{2n,2m} 
	\int_{-r}^r x^{2n}\,dx \int_{-r}^r y^{2m}\,dy \\
	&=& \sum_{2n+2m \le q} c_{2n,2m} \frac{2r^{2n+1}}{2n+1}\,
	\frac{2r^{2m+1}}{2m+1} \\
	&=& \sum_{2n+2m \le q} \frac{4c_{2n,2m}}{(2n+1)(2m+1)}\,r^{2n+2m+2}.
\eea
For an annulus with inner radius $\ri$ and outer radius $\ro$, the expected
number of counts is given by
\be
{\cal N}_\area = \sum_{2n+2m \le q} \frac{4c_{2n,2m}}{(2n+1)(2m+1)}\,
	\left(\ro^{2n+2m+2}-\ri^{2n+2m+2}\right)
\ee
Following \eq{B-define}, we divide by the area of the annulus to get the
count density 
\be
B_\area = \sum_{2n+2m \le q} \frac{c_{2n,2m}}{(2n+1)(2m+1)}\,
	\left(\frac{\ro^{2n+2m+2}-\ri^{2n+2m+2}}{\ro^2-\ri^2}\right).
	\label{eq-B-poly}
\ee
In particular, if we define
\be
	L_k = \sum_{j=0}^k \frac{k+1}{(2j+1)(2k-2j+1)}c_{2j,2k-2j}
\ee
and invoke the definition in \eq{define-Rk}, we can write
\be
	B_\area = \sum_{k=0}^{\qfloor} L_k R_k.
	\label{eq-LR-decomp-sq}
\ee
where $L_k$ depends only on the underlying polynomial, and $R_k$ depends only
on the geometry of the annulus.  Here $\qfloor$ is the ``floor'' of $q/2$;
that is, the largest integer less than or equal to $q/2$.
What is remarkable about this expression is that there are only $1+\qfloor$ 
variables (the $L_k$'s) which depend on the $(q+1)(q+2)/2$ polynomial
coefficients $c_{nm}$.  If our goal is to estimate $B_\area$ for some area
$\area$, then we only need to estimate $1+\qfloor$ separate coefficients.

\subsection{Average counts in a circular annulus}
It is fairly straightforward to adapt the above derivation to circular
annuli.  Writing \eq{mu} in polar coordinates, we get
\be
	{\cal N}_\area = \int_\area B(x,y) \,dx \,dy = 
	\int_{\ri}^{\ro} r\,dr\, \int_0^{2\pi} B(r\cos\theta,r\sin\theta)\,d\theta
\ee
Expanding $B(x,y)$ as a polynomial of degree $q$ we obtain
\be
	{\cal N}_\area = \sum_{n+m \le q} c_{nm} \int_{\ri}^{\ro} r\,r^{n+m}\,dr
	\int_0^{2\pi} \cos^n\theta \sin^m\theta\,d\theta
\ee
It is clear that $\int_0^{2\pi} \cos^n\theta \sin^m\theta\,d\theta$ vanishes
if either $n$ or $m$ are odd (and that it is strictly positive when both
$n$ and $m$ are even), so we can rewrite:
\be
	{\cal N}_\area = \sum_{2n+2m \le q} c_{2n,2m} \int_0^r r\,r^{2n+2m}\,dr 
	\int_0^{2\pi} \cos^{2n}\theta \sin^{2m}\theta\,d\theta
\ee
Let us define
\be	
	K_{nm} \equiv \frac{1}{2\pi}
	\int_0^{2\pi} \cos^{2n}\theta \sin^{2m}\theta\,d\theta.
\ee
One can show that $K_{nm} = {(2m-1)!!(2n-1)!!}/{2^{n+m}(n+m)!}$,
where $n!! = n(n-2)(n-4)\cdots 1$, 
but we don't actually need the closed form.
In terms of $K_{nm}$, we can write
\be
	{\cal N}_\area = \sum_{2n+2m \le q}  
	\frac{c_{2n,2m}(2\pi K_{nm})}{2n+2m+2}\, 
	\left( \ro^{2n+2m+2} - \ri^{2n+2m+2} \right)
\ee
Now, define
\be
	L_k = \sum_{i=0}^k K_{i,k-i}\,c_{2i,2k-2i}
\ee
so that	
\be
	{\cal N}_\area = \sum_{k=0}^{\qfloor} \frac{\pi L_k}{k+1} 
	\left(\ro^{2k+2}-\ri^{2k+2}\right)
\ee
and then combining \eq{B-taylor} and \eq{define-Rk}
\bea
	B_\area &=& {\cal N}_\area/{\rm area}(\area) \\
	&=& \sum_{k=0}^{\qfloor} \frac{\pi L_k}{k+1}\,\,
	\frac{\ro^{2k+2}-\ri^{2k+2}}{\pi\ro^2-\pi\ri^2} \\
	&=& \sum_{k=0}^{\qfloor} L_k R_k.
	\label{eq-LR-decomp}
\eea
where, again, the $1+\qfloor$ variables $L_k$ depend on the underlying
polynomial but not on the geometry of the annuli, and the $R_k$ 
depend only on the annuli.

\subsection{Using multiple concentric annuli to estimate source background}

The advantage of the decomposition of $B_\area$ into a sum of the form
given in \eq{LR-decomp-sq}, and again in \eq{LR-decomp}, can be seen when we try
to estimate the background count density in a central region $B_\src$
from the count densities $B_s$ estimated from each of the concentric
annuli.

Assume that the background can be well-modeled (at least in a region
containing all the annuli) by an order~$q$ polynomial, and that we
have $1+\qfloor$ concentric annuli surrounding the point source
candidate.

  Let $R_{ks}$ denote the $R_k$ value given in \eq{define-Rk}
for the $s$-th annulus.  This is essentially the $2k$-th moment
and is a purely geometrical property of the $s$-th annulus.  
Let $B_s$ be the average count density 
in that annulus.  Under these assumptions, 
the $1+\qfloor$ annuli all satisfy
\be
	B_s = \sum_{k=0}^{\qfloor} L_k R_{ks}
\ee
which is a linear system of $1+\qfloor$ equations in $1+\qfloor$ unknowns.
In particular, let ${\cal R}_{sk}$ denote the $s,k$ element of the
inverse of the $R_{ks}$ matrix.  That is, $\sum_k {\cal R}_{sk}R_{ks'} 
= \delta_{ss'}$.
Then $L_k = \sum_s B_s{\cal R}_{sk}$.  If the central source region 
has moments
$R_{k,{\rm src}}$, then we can write the count density in the central area as
\be
	B_\src = \sum_k L_kR_{k,{\rm src}} =
	\sum_k \sum_s B_s {\cal R}_{sk} R_{k,{\rm src}}
\ee
or, if we define
\be
	\beta_s \equiv \sum_k {\cal R}_{sk} R_{k,{\rm src}},
	\label{eq-beta-unbiased}
\ee
then
\be
	B_\src = \sum_s \beta_s B_s,
	\label{eq-gen-linear-combination}
\ee
is a simple linear combination of the $B_s$'s, and the coefficients $\beta_s$ 
depend only on the geometry of the annuli.  In particular, if we estimate
$B_s$ with $\hat B_s=N_s/A_s$, then we can estimate $B_\src$ with 
\be
	\hat B'_\src = \sum_s \beta_s N_s/A_s,
	\label{eq-gen-linear-combination-est}
\ee
where the prime indicates that this estimate of $B_\src$ is obtained from the
surrounding annuli and not from the direct $N_\src/A_\src$.
Note that the $\beta_s$'s are not necessarily positive; thus, it is possible
for \eq{gen-linear-combination-est} to produce a negative estimate of the
background $B_\src$.

\subsubsection{An example: two concentric annuli}

\begin{figure}
\epsscale{0.75}
\plotone{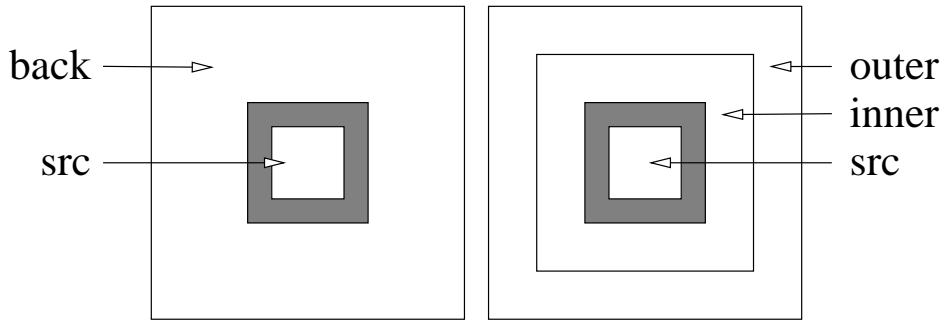}
\caption{The one-annulus method estimates 
background in the source region (src) with the observed
count density in the background annulus (back).
Using two annuli, the count density in the inner and outer
annuli are extrapolated to the center to provide an estimate
of the average background in the source region.}
\label{fig-annuli}
\end{figure}

Consider a source region $\area_{\rm src}$
surrounded by {\em two} concentric annuli: $\area_{\rm inner}$ and 
$\area_{\rm outer}$.  See \fig{annuli}.
If $B(x,y)$ can be accurately modeled as a cubic function of 
$(x,y)$ in the region containing the annuli, then we can write
\bea
	\bcen &=& L_0 + L_1\rcen^2 \nonumber \\
	\binn &=& L_0 + L_1\rinn^2 \label{eq-bvsr}\\
	\bout &=& L_0 + L_1\rout^2 \nonumber
\eea
where $L_0$ and $L_1$ are two scalars which
depend on the underlying cubic background function,
and $\rcen$, $\rinn$, and $\rout$ are the ``average radii'' of the center,
inner, and outer annuli.  This average is defined in terms of the second
moment for each of the three areas:
\be 
	r_\area^2 = R_{1,\area} = \frac{1}{2}
	\left(r_{i,\area}^2 + r_{o,\area}^2\right).
\ee
Here $r_{i,\area}$ is the inner radius, and $r_{o,\area}$ is the outer radius,
of the annulus $\area$.
We can now estimate the average
count density for the central area as a linear combination of the
count densities in each of the two surrounding annuli:
\be
   \bcen = \beta\binn + (1-\beta)\bout
	\label{eq-bcen-twoannuli}
\ee
with
\be
   \beta = \frac{\rout^2-\rcen^2}{\rout^2-\rinn^2}.
	\label{eq-beta-twoannuli} 
\ee 
We remark again that $\beta$ depends only on the geometry of the
the background annuli and the source region.  It does not depend on
the underlying background nonuniformity.  Thus, having set up a geometrical
configuration, one needs to compute $\beta$ only once.  We also remark
that since $r_\inn > r_\src$, we have $\beta>1$; this implies that
the coefficient of $\bout$ is always negative.
A ``typical''
value is $\beta=1.5$ (and $1-\beta=-0.5$), which corresponds to the case that 
both annuli have the same area and are much larger than the
hole in the inner annulus.

\begin{figure}
\epsscale{0.5}
\plotone{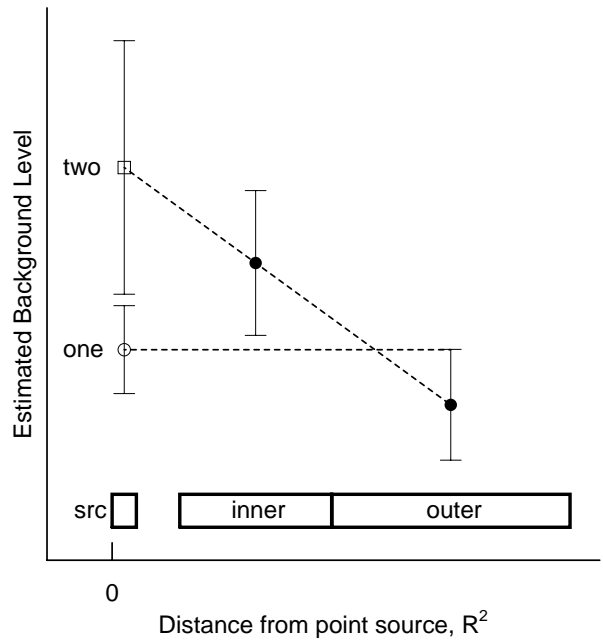}
\caption{The background in the source region (src) is estimated from
the counts per pixel in the inner and outer annuli.  The two-annulus
estimator (open square symbol) provides a direct extrapolation of the
outer and inner count densities to the source region.  The one-annulus
estimator (open circle symbol)
treats the inner and outer annuli as a single annulus, and
so computes an average of the inner and outer background estimates weighted
by their respective areas.  The error bars indicate
plus and minus one-sigma errors (where ``sigma'' is the square root of
the variance of the estimator) for each of the estimated background levels.
The two-annulus estimator has a much larger variance 
than the one-annulus estimate, but it is
unbiased even for cubic polynomial nonuniformities in the background.
}
\label{fig-bvsrr}
\end{figure}

It is clear from \eq{bvsr} that a plot of $B$ versus
$r^2$ will produce a straight line.  This provides a convenient visualization
(as seen in figure~\ref{fig-bvsrr}) of the ``linear extrapolation'' of 
the background from the outer and inner
rings to the central source region.

\subsubsection{Overdetermined case}

Although we need at least $1+\qfloor$ annuli to fit an order $q$ polynomial,
there is no reason not to use more than the minimum number of annuli.
For the $q=3$ case, we can imagine replacing the ``linear extrapolation''
in \fig{bvsrr} with a linear fit; if we do it this way, however, we have
to be careful about the Poisson bias~\cite{Wheaton95}.  In general, 
we are free to choose our coefficients $\beta_s$ however we
like so long as the following conditions are met:
\be
	R_{{\rm src},k} = \sum_s \beta_s R_{s,k}
\ee
for $k=0,\ldots,\qfloor$.  These conditions ensure that if the background
varies as an order-$q$ polynomial, then \eq{gen-linear-combination} will
hold, and we will have an unbiased estimator.
In later sections,
we will discuss different approaches
for optimizing the choice of the coefficients $\beta_s$.

It is useful to consider the limit of an infinite number of annuli,
and to take the continuum limit.  
The role of the coefficients $\beta_s$
is then 
played by a function $\beta(r)$ where $r$ is the radius of the infintesimal
annulus.\footnote{Actually, the $\beta(r)$ that we will use 
is not exactly the continuum analog of
$\beta_s$; it's more like $\beta(r) \sim \beta_s/A_s$.}  
The estimator for the average background in the source region is given by
an integral
\be
	\hat B'_\src =
        \int \beta(r) B(r) 2\pi r\,dr
	\label{eq-Bhat-continuum}
\ee
but this can be more usefully written as a sum over all the individual
counts (photons):
\be
	\hat B'_\src =
	\sum_{n=1}^N \beta(r_n) 
\ee
where $r_n$ is the distance\footnote{The distance is Euclidean 
($r=\sqrt{x^2+y^2}$) if we are
using circular annuli -- which we might as well do, if we have distances
available as floating point numbers.  The distance is given by the ``maximum''
metric  ($r = \max(x,y)$) if we are using square annuli.}
from the candidate point source location to
the position of the $n$-th photon.
The conditions to fit an order $q$ polynomial are obtained by setting
$B(r)=r^{2k}$ for $k=0,\ldots,\qfloor$,
and employing \eq{Bhat-continuum} to obtain
\be
	\int \beta(r) 2\pi r^{2k+1}\,dr = R_{{\rm src},k}.
	\label{eq-constraints-continuum}
\ee
Note that in the limit as the radius of the source
region goes to zero,
we have $R_{{\rm src},k} = \delta_{k,0}$.
In this limit, we are estimating the background
density at a point, instead of estimating the average background over a 
region.

Equations~(\ref{eq-Bhat-continuum},\ref{eq-constraints-continuum}) 
are written for circular annuli; essentially the
same result can be applied for square annuli, but with `$2\pi$'
(perimeter of unit circle) replaced by `8' (perimiter of unit-radius square).

The use of a continuum kernel for simultaneous point source detection
and background subtraction is also possible (\citeN{Damiani97} use 
wavelets for this purpose).  The optimal kernel can be constructed to
match the point spread function of the telescope and at the same time
be insensitive to polynomial background nonuniformity.

\section{Error (variance and bias) in background estimation}

In the limit as the background densities $B_s$ are precisely known,
the estimate of $B_\src$ given by
\eq{gen-linear-combination} will be precise.  In practice, however,
these background densities are estimated from a finite number of counts
in each of the annuli. 

If there are $N_s$ counts in an area $A_s$ in the $s$-th annulus, then
$\hat B_s = N_s/A_s$ is the estimated count density.  And since these
counts arise from a Poisson source, the average squared error in $\hat
B_s$ will be given by
\be
	E_s^2 =
	\langle ( \hat B_s - B_s )^2 \rangle = B_s/A_s.
	\label{eq-variance-singleterm}
\ee
Using this expression, we can write the averaged squared error 
for our estimate in \eq{gen-linear-combination-est}
of the count density in the source area as
\be
	E_\src^2 = 
	\langle ( \hat B'_\src - B_\src)^2\rangle =
	\sum_s \beta_s^2 E_s^2 = \sum_s \beta_s^2 B_s/A_s
	\label{eq-variance}
\ee
It is straightforward to extend this expression to the continuum limit:
\be
	E_{\src}^2 = \int
	\beta^2(r) B(r)\,2\pi r\,dr.
	\label{eq-variance-continuum}
\ee
where we define $B(r)\equiv\frac{1}{2\pi}\int_0^{2\pi} B(r,\theta)\,d\theta$.

The above equations express the {\em variance} in the estimate of
background count density at the source location; in general, there
will also be a {\em bias}.  If the background is well fit by a polynomial
of order $q$, then the bias is zero.
One way to estimate this bias is to assume that the background is well-fit
by an order $q'$ polynomial, with $q'\ge q$.  Then the 
bias in the order-$q'$ multi-annulus
estimator is zero, and the bias in the order-$q$ estimator will be given by the
difference between these two estimators.  In general, using smaller values
of $q$ will increase the bias, but at the same time reduce the variance.  
Determining the optimal order $q$ is therefore a trade-off between bias 
and variance.  (In our experience, however, 
the optimal $q$ is never larger than 3.)

\subsection{Uniform background}

For the moment, consider the case that the background is uniform
(so that $B_s=B$ for all $s$).  The bias in this case is zero,
and only the variance is relevant.
For the single annulus estimator, the variance is given by
\be
	V_1 = B/A
	\label{eq-varone}
\ee
where $A=A_\inn+A_\out$.  For the two annulus estimator, we have
\bea
	V_2 &=& \beta^2 B/A_\inn + (1-\beta)^2 B/A_\out = 
	\left(\frac{\beta^2}{\alpha} 
	+ \frac{(1-\beta)^2}{1-\alpha}\right) B/A \\
	&=& \left(1 + \frac{(\alpha-\beta)^2}{\alpha(1-\alpha)}\right) B/A
	\label{eq-vartwo}
\eea
where $\beta$ is given by \eq{beta-twoannuli}, and we have defined
\be
	\alpha = A_\inn/(A_\inn+A_\out).
	\label{eq-alphadefine}
\ee
It it clear from these expressions that 
$V_2$ can never be smaller than $V_1$; in fact, from the geometrical
constraint that the annuli and source region cannot overlap, it is
not too hard to show that $V_2/V_1$ is always larger than 
five.\footnote{%
The factor of five is achieved in the limit as the inner radius of 
the inner annulus goes to zero, 
and $A_\inn=A_\out$; we have $\beta=1.5$, $\alpha=0.5$ in this case, and
then from \eq{vartwo}, $V_2/V_1=5$.}

Note that \eq{vartwo} provides a criterion for 
choosing an optimal partition of a background
annulus into inner and outer annuli.  When the background is uniform,
it is straightforward to show that this optimum
partition occurs when the areas of the inner and outer annuli are equal.
In this case, $\alpha=0.5$, and
$V_2 = \left(1+(2\beta-1)^2\right)V_1.$

When the background is uniform, preference clearly goes to the
one-annulus estimator, since it has substantially smaller variance.
If we were to compute the variance for a three-annulus estimator which
fit an order $q=5$ polynomial to the background, we would find an even
larger variance.  On the other hand, using more than two annuli, but
sticking to $q=3$, will reduce the variance somewhat.  This is most
evident in the continuum limit.

In the continuum limit, there is a simple criterion for choosing the optimal
function $\beta(r)$: minimize the variance in \eq{variance-continuum}
while maintaining the constraints in \eq{constraints-continuum}.
Using the method of Lagrange multipliers, and some variational
calculus, one can show that the optimal
function is of the form
\be
	\beta(r) = \frac{\displaystyle\sum_{k=0}^{\qfloor} \lambda_k r^{2k}}%
	                {B(r)}.
\ee
where $\lambda_k$ are constants whose values are determined from the
conditions in \eq{constraints-continuum}
for $k=0,\ldots,\qfloor$.

\begin{figure}
\epsscale{0.5}
\plotone{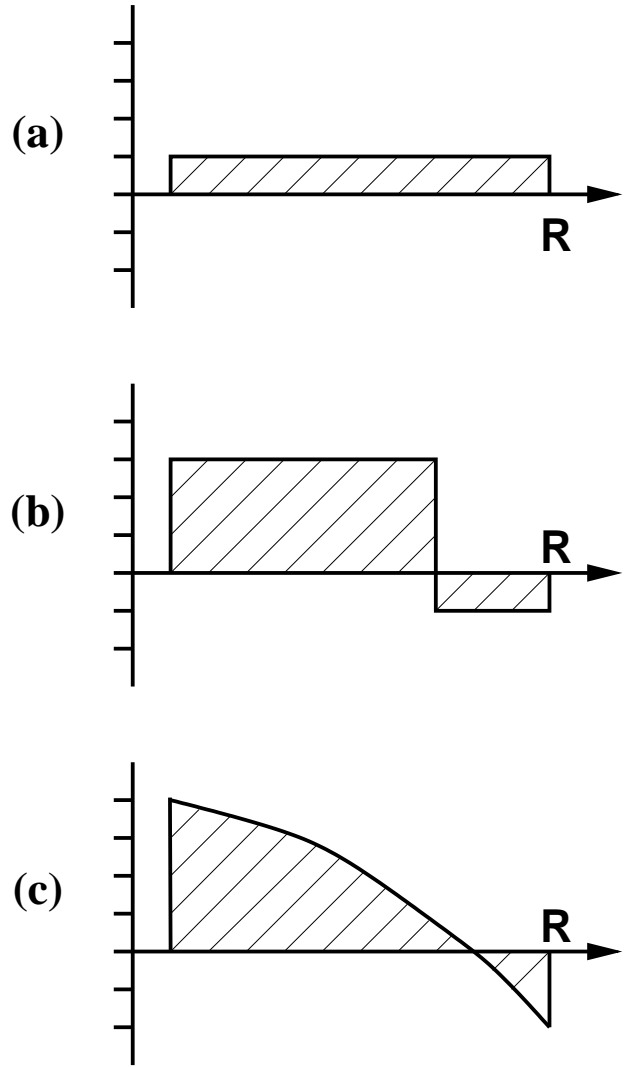}
\caption{Radial kernels for background estimation at the origin:
(a) Simple one-annulus estimator; if the background is uniform,
then this provides the optimal estimate.  (b) Two-annulus estimator; 
if the background has nonuniformities which can be modelled with
a cubic polynomial, then this estimator is unbiased.
(c) The continuum limit of a multi-annulus estimator is a kernel function 
which varies with radius.  One can derive an optimal kernel function
-- see \protect{\eq{optimal-continuum}} -- which produces the least
estimator variance while being constrained to being unbiased
for cubic nonuniformities.}
\label{fig-ann-kern}
\end{figure}

As an example, consider the case that $B(r)$ is constant, that $q=3$,
and that the source region and annulus inner
radius are both much smaller than the annulus outer radius.  In that case,
we have $\beta(r)=a+br^2$.  The constraints in \eq{constraints-continuum}
become
\bea
	\int_0^R \beta(r)\,2\pi r\,dr &=& 
	2\pi\left(\frac{aR^2}{2} + \frac{bR^4}{4}\right) = 1 \\
	\int_0^R \beta(r)\,2\pi r^3\,dr &=&
	2\pi\left(\frac{aR^4}{4} + \frac{bR^6}{6}\right) = 0 
\eea
and from these, we obtain
\be
	\beta(r) = \frac{4}{\pi R^2} - \frac{6}{\pi R^4}\,r^2.
	\label{eq-optimal-continuum}
\ee
as the optimal coefficient function.  
The variance of this optimal continuum estimator
is
\bea
	V_C &=& \int_0^R \beta^2(r) B(r)\,2\pi r\,dr \\
            &=& 2\pi B \int_0^R \left[\frac{4}{\pi R^2} 
	- \frac{6}{\pi R^4}\,r^2\right]^2\,r\,dr
	    = \frac{4B}{\pi R^2} = 4B/A.
\eea
We see that this value is four times larger than the single-annulus
variance, but twenty percent smaller than the equivalent 
two-annulus variance.  (See \fig{ann-kern}.)

\subsection{Nonuniform background}
While it is useful to understand the performance of the various
estimators in the specical case of uniform background, the whole
purpose of these multi-annulus estimators is to improve the
characterization of nonuniform backgrounds.  With nonuniform backgrounds,
the bias is in general nonzero, and so bias/variance tradeoffs 
need to be considered.

For a two annulus estimator, if the background is nonuniform, 
then $B_\inn \ne B_\out$ in general.  Let
\be
	B=\alpha B_\inn + (1-\alpha)B_\out
\ee
be the average background,\footnote{$B$ is also the source background that would
be estimated by the one-annulus estimator.  Recall $\alpha$ was defined in
\eq{alphadefine} in terms of the ratio of areas of the inner and outer annuli.} 
and let
\be
	\delta = \frac{B_\inn-B_\out}{B}
	\label{eq-delta-define}
\ee
be a measure of nonuniformity which is restricted to 
$-1\le\delta\le 1$.  Here, positive $\delta$ corresponds to a ``peak'' at the
source location, and 
negative $\delta$ to a valley.
Note that we can express arbitrary linear combinations
of $B_\inn$ and $B_\out$ in terms of the average background and this
nonuniformity measure.
\be
	uB_\inn + vB_\out = (u+v)B + ((1-\alpha)u-\alpha v)\delta B
	\label{eq-lincombinerule}
\ee

If the background nonuniformity 
is well modeled by a cubic polynomial, then the bias in the two-annulus
estimator will be zero;  this means that the bias for the one annulus
estimator is given by the difference between the one and two annulus
estimators.
\bea
	\langle \hat B'_\src  - B_\src \rangle &=&
	B -
	(\beta B_\inn + (1-\beta)B_\out) \nonumber \\ &=&
	(\alpha B_\inn + (1-\alpha) B_\out) -
	(\beta B_\inn + (1-\beta)B_\out) \nonumber \\ &=&
	(\alpha-\beta)\delta B 
\eea
We can now write the total relative squared error (variance plus 
squared bias divided by $B^2$) for the one-annulus estimator:
\be
	T_1^2 = \frac{1}{{\cal N}} + (\alpha-\beta)^2\delta^2
\ee
where ${\cal N}=AB$ is the expected
number of counts in the background annulus.

For the two-annulus estimator, the bias is zero but the variance is
given by
\be
	V_2 = \beta^2 B_\inn/A_\inn + (1-\beta)^2 B_\out/A_\out
	\label{eq-twoannvariance-raw}
\ee
Using \eq{lincombinerule} to expand the above sum, 
we can write an expression 
for the total relative squared error:
\be
	T_2^2 = V_2/B^2 =
	\left\{
	\left(\frac{\beta^2}{\alpha} + \frac{(1-\beta)^2}{1-\alpha}\right) +
	\left(\frac{\beta^2(1-\alpha)}{\alpha}
	    - \frac{(1-\beta)^2\alpha}{1-\alpha}\right)\delta
	\right\}\,\frac{1}{{\cal N}}.
	\label{eq-twoannvariance}
\ee
If $|\delta|\ll 1$, we can write more simply
\be
	T_2^2 = \left(1 + \frac{(\alpha-\beta)^2}{\alpha(1-\alpha)}\right)\,
	\frac{1}{{\cal N}}.
\ee
One prefers the two-annulus method if $T_2^2 < T_1^2$, and this happens
when 
\be
	{\cal N}\delta^2 > \frac{1}{\alpha(1-\alpha)}.
\ee
The two-annulus estimator is preferred when the bias is large ($\delta$ large) 
and the variance is small (${\cal N}$ large).  If the background annuli cover
a large area, then both of these conditions are more likely to hold.

When the background is nonuniform, the optimal partition of the 
background annulus into two annuli is no longer into equal areas.
In general, local peaks ($\delta>0$) prefer a larger inner annulus,
and local valleys ($\delta<0$) prefer a larger outer annulus.  However,
the gain for mild nonlinearities is not substantial, and since one usually
prefers to use the same-sized annuli for the entire map, we recommend 
$A_\out \approx A_\inn$ as a general rule of thumb.

\subsection{Combination of one- and two-annulus methods}

In comparing the one- and two-annulus methods, it is useful to recognize
that both methods provide an estimate of $B_\src$ as a linear combination
of 
$\hat B_\inn$ and $\hat B_\out$, with the only difference being the
choice of coefficients:
\be
	\hat B'_\src = \gamma \hat B_\inn + (1-\gamma) \hat B_\out.
	\label{eq-gammadefine}
\ee
Here, $\gamma = \alpha$ with $\alpha$ defined in \eq{alphadefine} 
produces the one-annulus
estimator, and $\gamma=\beta$ with $\beta$ defined in \eq{beta-twoannuli}
gives the two-annulus estimator.  Rather than try to make a dichotomous
choice between the one- and two- annulus estimators, we can instead ask 
about the optimal choice of $\gamma$.  In principle, it is straightforward
to derive the optimum; we have
\be
	T_\gamma^2 =
	\left\{
	\left(1 + \frac{(\alpha-\gamma)^2}{\alpha(1-\alpha)}\right) +
	\left(\frac{\gamma^2(1-\alpha)}{\alpha}
	    - \frac{(1-\gamma)^2\alpha}{1-\alpha}\right)\delta
	\right\}\,\frac{1}{{\cal N}} +
	 (\gamma-\beta)^2\delta^2,
\ee
so for a given ${\cal N}$ and $\delta$, we can take a derivative and set it to zero.
Using the $\delta\ll 1$ approximation, this produces:
\be
	\gamma_{\mbox{\footnotesize optimal}} =
	\frac{\alpha + \zeta\beta}
	     {1      + \zeta}
	\label{eq-gamma-adaptive}
\ee
where $\zeta= {\cal N}\delta^2\alpha(1-\alpha)$.
Note that this expression for the optimal $\gamma$ depends on 
${\cal N}$ and $\delta$, which will
vary over different parts of the map.  There are two approaches we can take
in this situation.  One is to develop an ``adaptive'' formula 
for $\gamma$ that varies with
position on the map; its value will depend on local properties of the map,
which will have to be estimated at each position.  
A second approach is to find a single best $\gamma$ for the
entire map.  

\subsubsection{Adaptive linear combination}
  The simplest adaptive linear combination of $B_\inn$ and $B_\out$ 
uses the expression for $\gamma$ in \eq{gamma-adaptive}, but this 
requires an estimate of the nonuniformity $\delta$.
A natural estimator (following \eq{delta-define}) takes
$\hat\delta = (\hat B_\out - \hat B_\inn)/\hat B$, but 
this will generally overestimate $\delta^2$, which will make
$\gamma$ larger than optimal.  Roughly,
\be
	\langle(\hat\delta)^2\rangle \approx
	\delta^2 + \frac{1}{\alpha(1-\alpha){\cal N}}
\ee
which suggests using $\widehat{\delta^2}=(\hat\delta)^2 - 
\frac{1}{\alpha(1-\alpha){\cal N}}$; in other words, use
\eq{gamma-adaptive} with
\be
	\zeta = \max\left\{0,\,
	\frac{A_\inn A_\out(\hat B_\out - \hat B_\inn)^2}%
	     {(A_\inn+A_\out)\hat B} - 1\right\}.
\ee

\subsubsection{Average best linear combination}
\label{sect-avgbest}

Although the adaptive estimate of $\gamma$ in the previous section should in
principle be the most accurate (at least for high enough count densities),  
we have obtained better estimates using a single ``average best'' $\gamma$.
Let $\hat B_\src$ denote the estimated counts per
pixel in the source area using the actual counts in the source area:
\be
	\hat B_\src = \frac{N_\src}{A_\src}
\ee
and let $\hat B'_\src$ be the estimated counts per pixel in the source area
as estimated from the counts in the background annuli.
If there is not a point source at the candidate location, then
both estimates should roughly agree.  Let
\be
	\Delta\hat B_\src = \hat B_\src - \hat B'_\src
	\label{eq-define-deltaB}
\ee
be the extent to which they do not agree.  Note that $\Delta\hat B_\src$
is a difference of two {\em estimators}; it is a quantity that can be computed
directly from a map without any knowledge of the actual background $B_\src$.  

This empirical estimate of estimation error can be 
combined with \eq{gammadefine} to produce
a tool for choosing a
single ``average best'' parameter $\gamma$. Note that
\be
\Delta\hat B_\src = \hat B_\src - \left( \gamma \hat B_\inn 
	+ (1-\gamma)\hat B_\out\right)
\ee
and
\be
	\langle (\Delta\hat B_\src)^2 \rangle =
	\langle (\hat B_\src - \hat B_\out)^2 \rangle
	- 2\gamma \langle 
	(\hat B_\src - \hat B_\out)(\hat B_\inn-\hat B_\out) \rangle
	+ \gamma^2 \langle (\hat B_\inn - \hat B_\out)^2 \rangle
\ee
and this is a simple quadratic equation whose minimum occurs at
\be
	\gamma_{\mbox{\scriptsize optimal}} =
	\frac{\langle(\hat B_\src - \hat B_\out)(\hat B_\inn-\hat B_\out) \rangle}%
	{\langle (\hat B_\inn - \hat B_\out)^2 \rangle}.
	\label{eq-gamma-avebest}
\ee
Since this expression uses $\hat B_\src$ which is estimated by counting 
photons in the source region, this estimate of 
$\gamma_{\mbox{\scriptsize optimal}}$ will only be useful in a map that
does not have a lot of real sources.  If there are a few bright sources,
these should
be masked off in the estimate of $\gamma_{\mbox{\scriptsize optimal}}$;
if there are many sources, 
then one must iteratively find, fit, and subtract off the real sources.
The difficulties involved in this source-confused regime are beyond the 
scope of this article.

\section{Applications of multiple annuli}

\begin{figure}
\plottwo{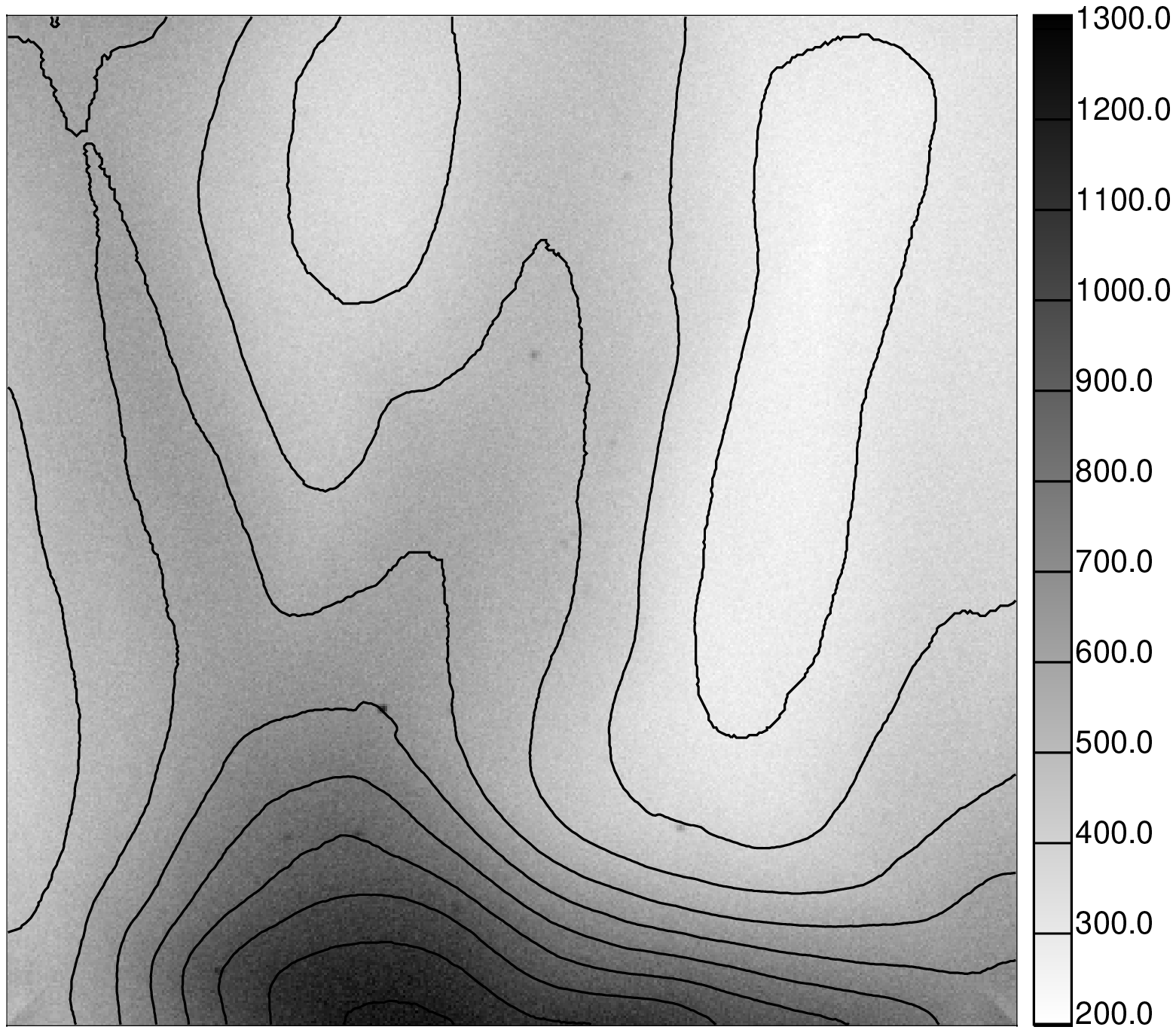}{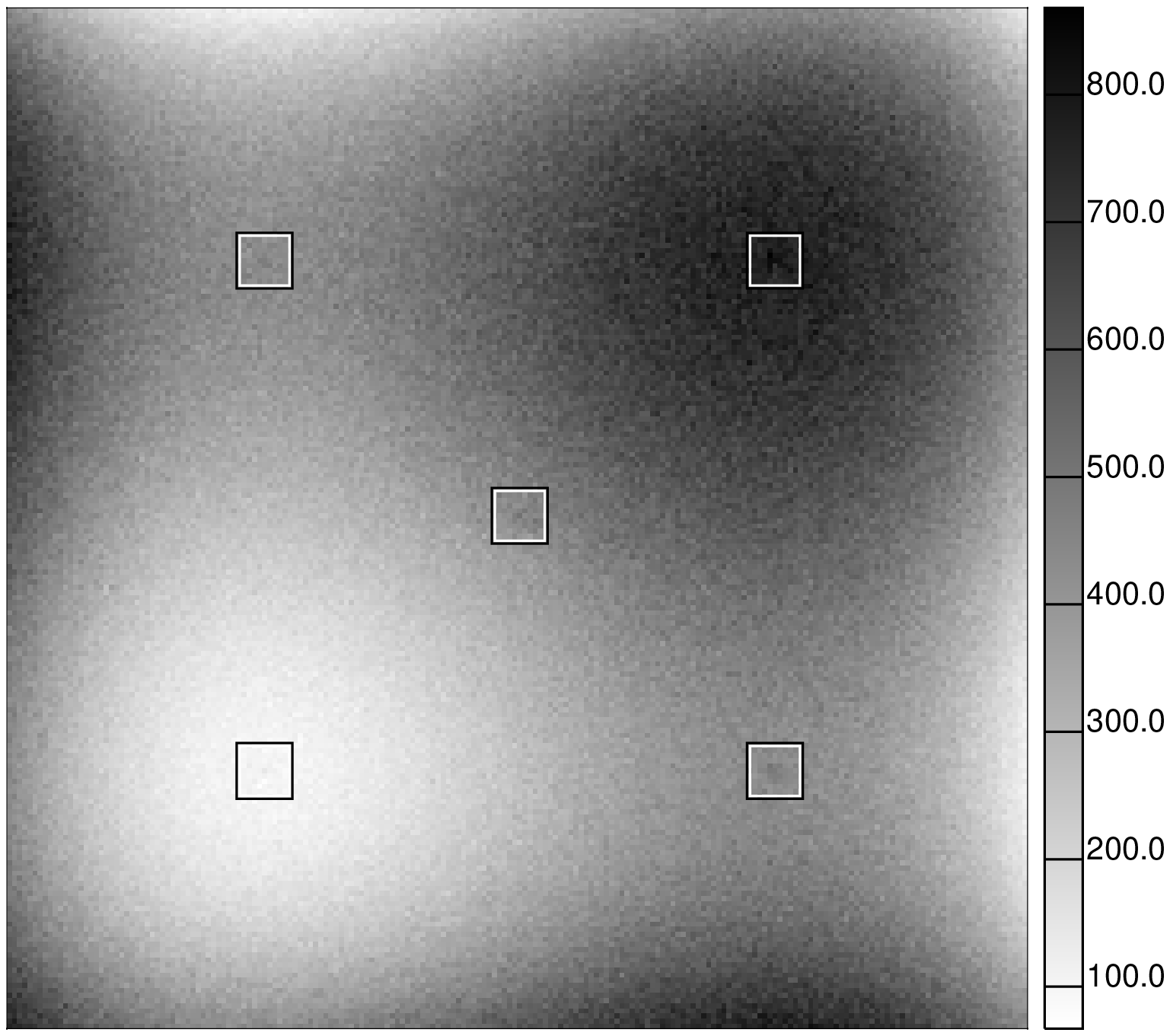}
\caption{
{\bf (a)} This data set constitutes one face, centered at $90^o$ Right
Ascension, and $0^o$ Declination, of the quadrilateralized
spherical cube projection of a full-sky map generated by
co-adding 49 months of raw count data from one of
the six ALEXIS telescopes.  The telescope we chose (1B), has a high
signal to noise ratio compared to the other telescopes, and
detects extreme ultraviolet photons in a narrow band centered at 172\AA.
The brightest object in this view is Sirius B,
the white dwarf companion of 
the visibly brightest star in the sky; it is in the lower left quadrant,
but is not easy to see before subtracting the background.
The contour lines indicate the irregular, but relatively smooth, 
spatial nonuniformity of the background; most of this
nonuniformity is due to the uneven exposure of the scanning telescope.
To avoid edge effects, the 360$\times$360 
map has been extended by 50 pixels in each direction by including
pixels from the four adjacent faces, and the
corners have been filled in by reflecting pixels across the edges.
The padded pixels are used only for background annuli; the point sources
are only detected in the original quad-cube face.
{\bf (b)} An artificial data set was generated to have an exactly cubic
polynomial
background.  Five artificial point sources were added at locations on the
map indicated by the enclosing squares.  All five have a source strength
that would be significant at 4 sigmas, if the background were precisely
known.  The points are arranged like the five dots on a die, and
each ``point'' is in fact a
3x3 source region with weight 0.2 in the center and 0.1 at each of 
the eight surrounding pixels.  In this 200$\times$200 map, the outer
30 pixels are reserved for ``padding;'' points are detected only in
the interior.
}
\label{fig-image}
\end{figure}

  We will illustrate a number of practical uses for multiple annulus
background estimation 
on a real ALEXIS data set (\fig{image}a) and on an artificial
data set (\fig{image}b) which has a cubic polynomial background and
five artifical point sources.  Four different applications of
multiple annulus methods will be provided: estimating
background level; detecting point sources; generating surrogate maps
for estimating false alarm rates;
and quantifying background nonuniformity.

\subsection{Estimating background level}
\label{sect-ebei}
In comparing various geometries of
single and multiple annulus estimators, it is useful to have an index
that measures the accuracy of these estimators directly from the data.
In Section~\ref{sect-avgbest}, we introduced an expression $\Delta\hat
B_\src=\hat B_\src - \hat B'_\src$ which describes the difference
between two estimates of the background in a source region.  The first
($\hat B_\src$) is a direct estimate from counts in the source region and the
second ($\hat B'_\src$) is an indirect estimate obtained from annular
regions that do not include the source region.  In the absence of a
real point source, we can evaluate the quality of the indirect annular
estimate by comparing it to the direct estimate.

Although we can compute $\Delta\hat B_\src$ directly from the data,
it is useful to note that we can write this as a difference of differences:
\be
	\Delta\hat B_\src = 
	(\hat B_\src - B_\src) - (\hat B'_\src-B_\src)
\ee

where $B_\src$ is the true (but unknown) background.
This suggests that there are two contributions to the total variance
in $(\Delta\hat B_\src)^2$: a source fluctuation and a background
estimation error.
The background estimation error itself arises from two causes: the
actual bias arising from the assumption that the background is a
low-order polynomial, and the Poisson fluctuations in the counts in
the annular regions.  Both of these are independent of the source
fluctuation error (which is due only to the Poisson fluctuation in the
counts in the source region), so these two contributions to the variance
(source fluctuation and background estimation error) are
independent, and we can write:
\be
	\langle(\Delta\hat B_\src)^2\rangle =
	\langle(\hat B_\src - B_\src)^2\rangle +
	\langle(\hat B'_\src-B_\src)^2\rangle.
\ee
with equality holding when $\langle\cdot\rangle$ denotes a true ensemble
average.  We can
furthermore estimate the first term (the source fluctuation error) from
the result in \eq{variance-singleterm}:
\be
	\langle (\hat B_\src - B_\src)^2 \rangle 
	= B_\src/A_\src \approx \hat B_\src/A_\src.
\ee
This gives an expression for the background estimation error:
\be
	\langle(\hat B'_\src - B_\src)^2\rangle \approx
	\langle (\Delta\hat B_\src)^2 \rangle
	- \hat B_\src/A_\src.
\ee
Dividing by the source fluctuation error, we
can obtain a directly computable dimensionless quantity
that we call the ``empirical background error index'':
\be
	{\cal E} = 
	\frac{\langle(\Delta\hat B_\src)^2\rangle}%
	     {\langle\hat B_\src\rangle /A_\src} - 1
	\label{eq-eei}
\ee
which is essentially the ratio of the average background estimation error to the
average source fluctuation error.
A smaller value is better, though
there is arguably a point of diminishing returns in obtaining a value 
very much less than one.  

\begin{figure}
\plotone{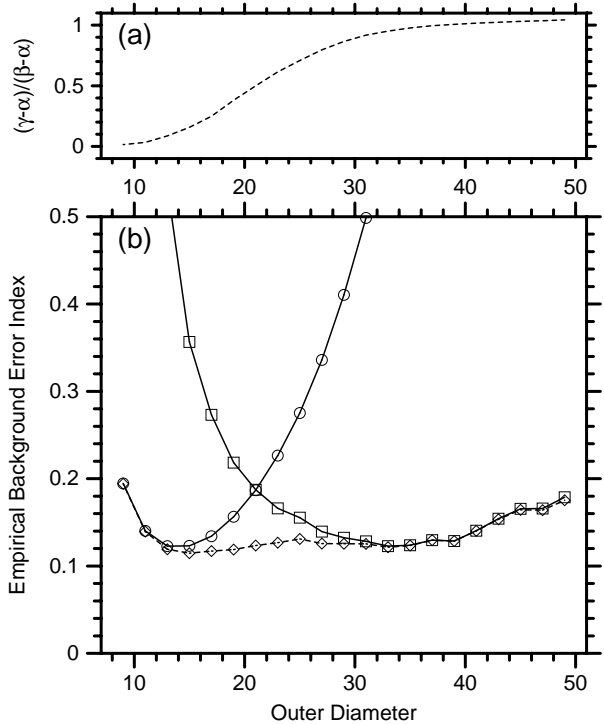}
\caption{
The Empirical Background Error Index for the ALEXIS map shown in 
\fig{image}a
is plotted against outer annulus diameter for the simple one-annulus
estimator (circles), for the two-annulus estimator in
\eq{bcen-twoannuli} (squares), for the ``average best'' linear
combination of one- and two- annulus estimators (dotted line, with
diamonds).  A smaller number indicates a more accurate background
estimation.  The annuli were square with odd-integer diameters; the
source region was 3x3 pixels, and the hole was 5x5.  For the two annulus
estimators, the annulus was partitioned to provide approximately equal
areas for the inner and outer annulus.
The dotted line in the upper panel shows
the coefficient of the best linear combination, rescaled so that
a value of zero corresponds
to the one-annulus estimator, and a value of one to the two-annulus
estimator.  For small annuli, the one-annulus estimator is preferred,
but for larger annuli, the two-annulus estimator is superior.
As the annulus diameter increases
further, however, the approximation of the background as a cubic polynomial
surface breaks down, and the two-annulus estimator gets worse.  At their
best annular diameters, the one-annulus and the two-annulus estimators 
exhibit virtually identical performance for this data set.}
\label{fig-eei-alexis}
\end{figure}

\begin{figure}
\plotone{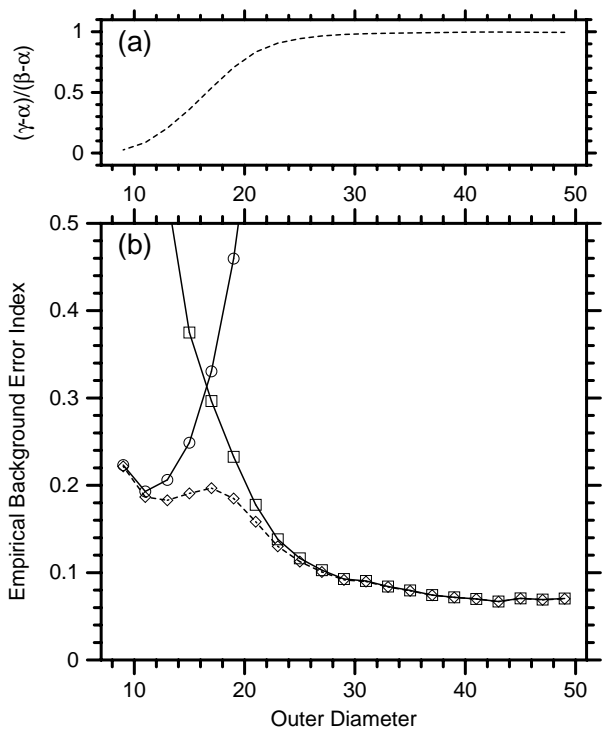}
\caption{
The Empirical Background Error Index for the artifical cubic background
map shown in \fig{image}b is plotted in just
the same way as shown in figure~\ref{fig-eei-alexis}.  Again, the one-annulus
estimator is preferred if the annulus is small, but the two-annulus estimator
is better as the annulus size grows.  In this
example, since the background is precisely cubic over the entire map, the
index decreases essentially monotonically with increasing annulus size.}
\label{fig-eei-cubic}
\end{figure}

In \fig{eei-alexis} (resp. \fig{eei-cubic}), we compare the empirical
background error index for one- and two-annulus estimators as a
function of outer annulus diameter for the ALEXIS (resp. artificial
cubic background) map.  For small annuli, the bias is smaller but the
variance is larger, and preference goes to the one-annulus estimator
which minimizes the variance.  For larger annuli, the bias is larger
but the variance is smaller, and preference goes to the two-annulus
estimator, which minimizes bias at the expense of variance.  For
intermediate-sized annuli, the optimal estimator is a linear
combination of the one and two annulus estimator.  These are only
qualitative trends; quantitative values ({\it e.g.,} optimal linear
coefficients, or optimal annulus size) depend on the background.  What
the empirical background error index provides is a way to estimate
these quantitative values directly from the raw data.

\subsection{Detecting point sources}

Our main motivation in attempting to more accurately estimate the
background is that this provides a more sensitive point source
detections -- in particular, it permits us to detect weaker real
sources without increasing the false detection rate.

When the background is known exactly, it is straightforward to assess
the significance of a given point source candidate.  If $N_\src$
counts are observed in the source region, and the background $B_\src$
is known exactly, then $S = N_\src - A_\src B_\src$ estimates the
source strength.  Under the null hypothesis, the
quantity $N_\src$ will be Poission distributed with mean $A_\src
B_\src$; the variance of $N_\src$ (and therefore of $S$) will also be
$A_\src B_\src$, so the ``signal to noise ratio'' will be given by 
\be
	S/N = \frac{S}{\sqrt{{\rm Var}(S)}} = \frac{N_\src - A_\src
	B_\src}{\sqrt{A_\src B_\src}}.  
\ee 
Under the null, $S/N$ will have
mean zero and variance one.\footnote{Note that we do not use
$\sqrt{N_\src}$ in the denominator, as some authors recommend. If we
did, then $S/N$ would not have mean zero and
variance one under the null hypothesis.}  If the count rate is high,
then $S/N$ will furthermore be distributed as a Gaussian, and
therefore it makes sense to think of the $S/N$ as a number of ``sigmas
of significance'' for a given point source detection.  For small
numbers of counts, the Gaussian approximation becomes inaccurate, and the
statistic $S/N$ is not sufficient to characterize the significance of
the point source detection. In that case, however, an exact test,
using Poisson statistics, is straightforward to derive. 
\citeN{Gehrels86}, for instance, provides tables and useful approximations 
for the Poisson formulas that arise in this context.
\citeN{Zepka94} argue for using a histogram to characterize the background;
this avoids assumptions of Gaussian or even Poisson statistics, but requires 
that that the
background be uniform over a large enough region that a good histogram
can be acquired.

But our interest is in the case that the background is not
known, and must be estimated by counts in surrounding annuli.  If the
background is estimated by $\hat B'_\src$, then the signal is given by
$S = N_\src - A_\src \hat B'_\src$, and the ``noise'' has an extra
contribution due to the variance of the estimator.  In particular, for
the multiple-annulus estimators, we can write
\be
	S = N_\src - A_\src\sum_s \beta_s N_s/A_s
	\label{eq-signal-multiannulus}
\ee
and the variance is given by
\be
	{\rm Var}(S) = A_\src B_\src + A_\src^2\sum_s \beta_s^2 B_s/A_s
	\label{eq-noise-multiannulus}
\ee
where $N_s$ (resp. $A_s$) is the counts (resp. area) in the $s$'th
annulus.  A more sensitive statistic would employ a matched filter
(see \citeN{Vikhlinin95} for a more extensive exposition on the use
of a matched filter for source detection, a flat annulus for
background subtraction, and Monte-Carlo tests for calibration).

The more precisely we can estimate the background,
the more sensitive is our test for point source candidates; in 
particular, 
anything that can be done to reduce this variance ({\it e.g.,} 
using background annuli with large area $A_s$) will
increase the statistical significance of any real sources, without altering the
false alarm rate.
This is a tangible motivation for estimating the background as precisely
as possible.

However, exact values of $B_\src$ and $B_s$ are not known in general,
so the variance itself must be estimated.  For the one-annulus
estimator, a number of authors have shown how this should be done.  In
particular, \citeN{Li83} argue (and demonstrate numerically as well)
that one should use the null hypothesis that the background is
uniform, and estimate $\hat B=(N_\src+N_\back)/(A_\src+A_\back)$.  That is,
\be	
	{\rm Var}(S) \approx A_\src \hat B + A^2_\src\hat B/A_\back \\
	             = (A_\src/A_\back)(N_\src+N_\back)
\ee
and
\be
	S/N = \frac{N_\src - (A_\src/A_\back)N_\back}%
	     {\sqrt{(A_\src/A_\back)(N_\src+N_\back)}}.
\ee
This statistic will have mean zero and variance one as long as the background
is flat ($B_\src = B_\back$) and there are no point sources.  It is only
approximately Gaussian, however, and the approximation becomes poorer as
the number of counts becomes small.  \citeN{Li83} also
provide a much more complicated estimate of S/N which is more accurately
Gaussian with lower counts.  
This more complicated expression can be further extended to
account for weighted counts~\cite{A_Theiler_ADASS_97}.  
The Cash
statistic~\cite{Cash79,1989ApJ...342.1207N}
provides a good tool for parameter estimation and confidence intervals 
even when the number of
counts in individual bins is small, but it 
is not really designed for point source detection.  \citeN{Damiani97}
employ a wavelet-based point source detector, essentially a continuum
circular source kernel and annular background, and provide an empirical
correction to Gaussian statistics with coefficients determined 
from a set of Monte-Carlo runs.

In \citeN{Lampton94}, the essentially binomial character of the counts
in the source and background regions is exploited to produce an exact
expression for statistical significance of counts in a source and
background region.  An extension of these exact results to nested
annuli is described in \citeN{A_Theiler_SCMA_97}.  Interestingly,
Lampton's formula is equivalent to one derived earlier in
\citeN{Alexandreas93}, but the earlier derivation is based on an
entirely different approach.  

But all of these exact results and corrections to Gaussian statistics
assume a uniform background.  For the nonuniform background, at least
for the purposes of this study, we will employ the implicit Gaussian
assumption in our use of signal to noise (or ``sigmas'') to
characterize significance.  However, in the next section, we will
describe an empirical approach for calibrating these sigmas to actual
significance.

The ``signal'' and ``noise'' for the multiple-annulus estimators are
given in \eq{signal-multiannulus} and \eq{noise-multiannulus}.  
As with the one-annulus estimator, the signal depends only on the 
counts in and the areas of the source region and background annuli,
but the noise depends on the true background levels.  To keep things
simple, we will follow the approach of \citeN{Li83}, and use
$\hat B=N_\total/A_\total$ 
where $N_\total=N_\src+\sum_s N_s$ and $A_\total=A_\src+\sum_s A_s$;
then 
\be
	{\rm Var}(S) = N_\total(A_\src/A_\total)
	\left(1 + A_\src\sum_s \beta_s^2/A_s\right)
\ee
and we measure significance with
\be
	S/N = \frac{N_\src - A_\src\sum_s \beta_s N_s/A_s}%
	{\sqrt{N_\total(A_\src/A_\total)
	\left(1 + A_\src\sum_s \beta_s^2/A_s\right)}}.
	\label{eq-sigmas}
\ee

\begin{figure}
\plottwo{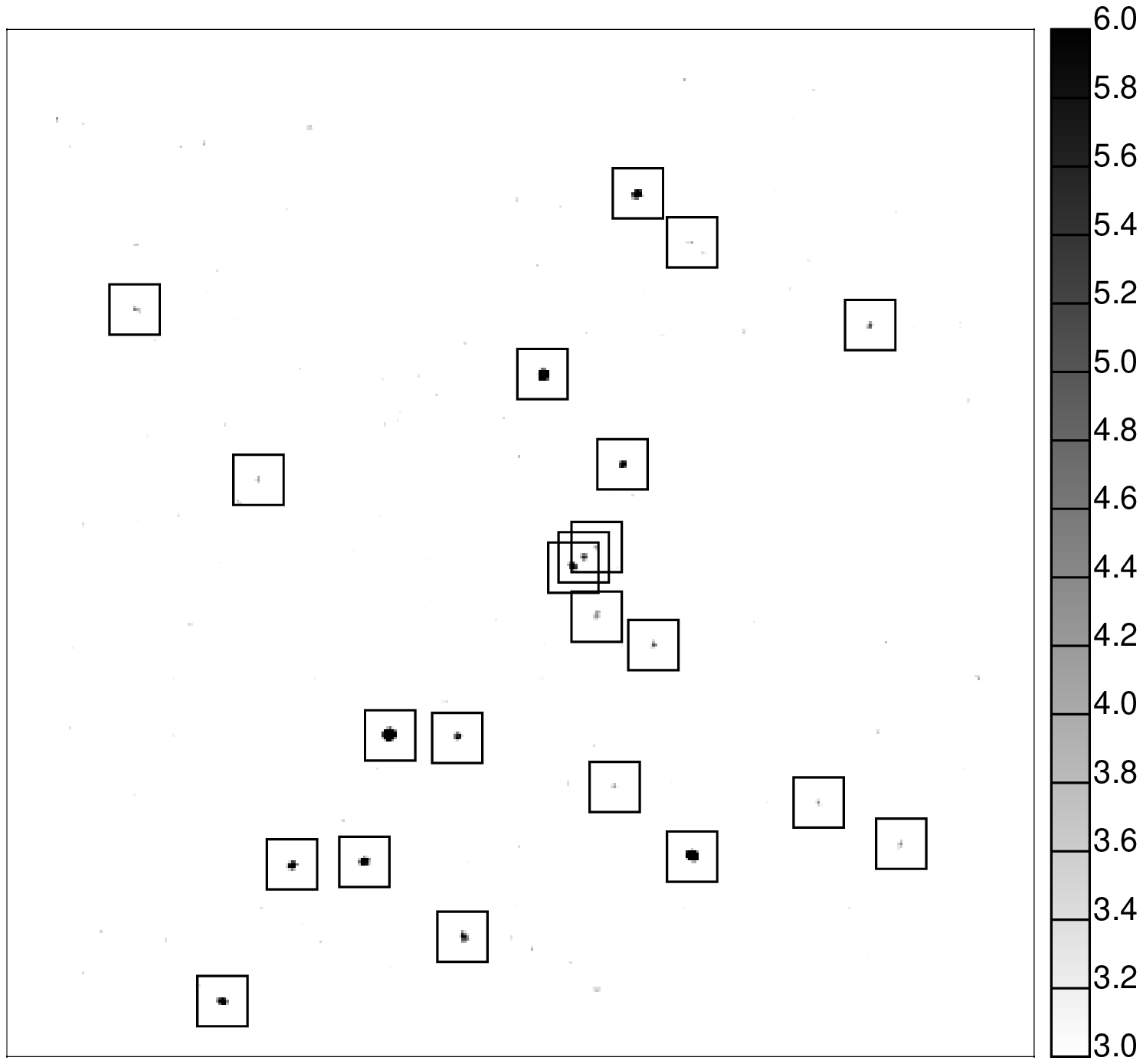}{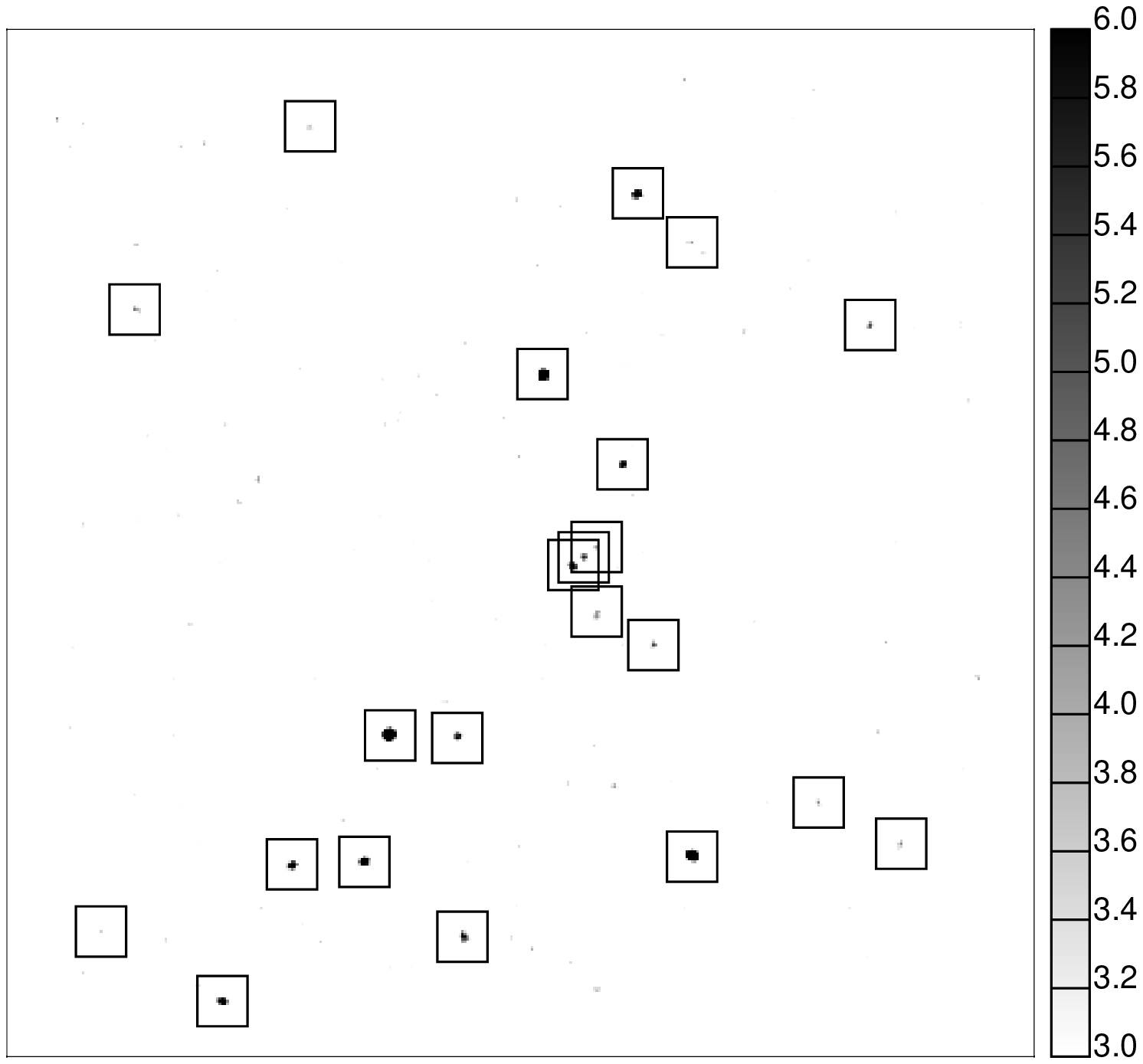}
\caption{Significance map
for the ALEXIS map shown in \fig{image}a, using
{\bf (a)} a one-annulus detector with a 13x13 pixel background, and
{\bf (b)} a two-annulus detector with a 25x25 inner annulus and a 35x35
outer annulus.  In both cases, the source region was 3x3 and the annular
hole was 5x5.  The significance is in units of ``sigmas'', or more
specifically, the $S/N$ statistic of {\protect\eq{sigmas}}.
Only significances above three sigmas are shown, and point source
detections with $S/N>4$ are indicated with squares.  (Those with
$S/N>5$ are also listed in Table 1.)
A number of known white dwarfs and bright O and B starts are 
visible in this map, including the distinctive belt of Orion, which
is near the center of this map.}
\label{fig-alexis-amap}
\end{figure}

In \fig{alexis-amap}, we show maps of the significance $S/N$ for the
ALEXIS data, as estimated using one- and two-annulus methods. 
A number of bright O and B stars and previously identified bright
extreme ultraviolet sources~\cite{EUVE2Catalog} are readily
identifiable, as well as others that may be unidentified transient
sources or false alarms; a more detailed study of these detections
will be reported in a later paper. 
Table~\ref{table-sigmas} shows all
sources that are detected at the four sigma level using one of the two
methods.  For those significant at the five sigma level,
identifications are provided as well.  For these identified sources,
there is a fairly close correspondence between the significance
level of the two methods, though the two-annulus detector is usually the
more significant.

\begin{figure}
\plottwo{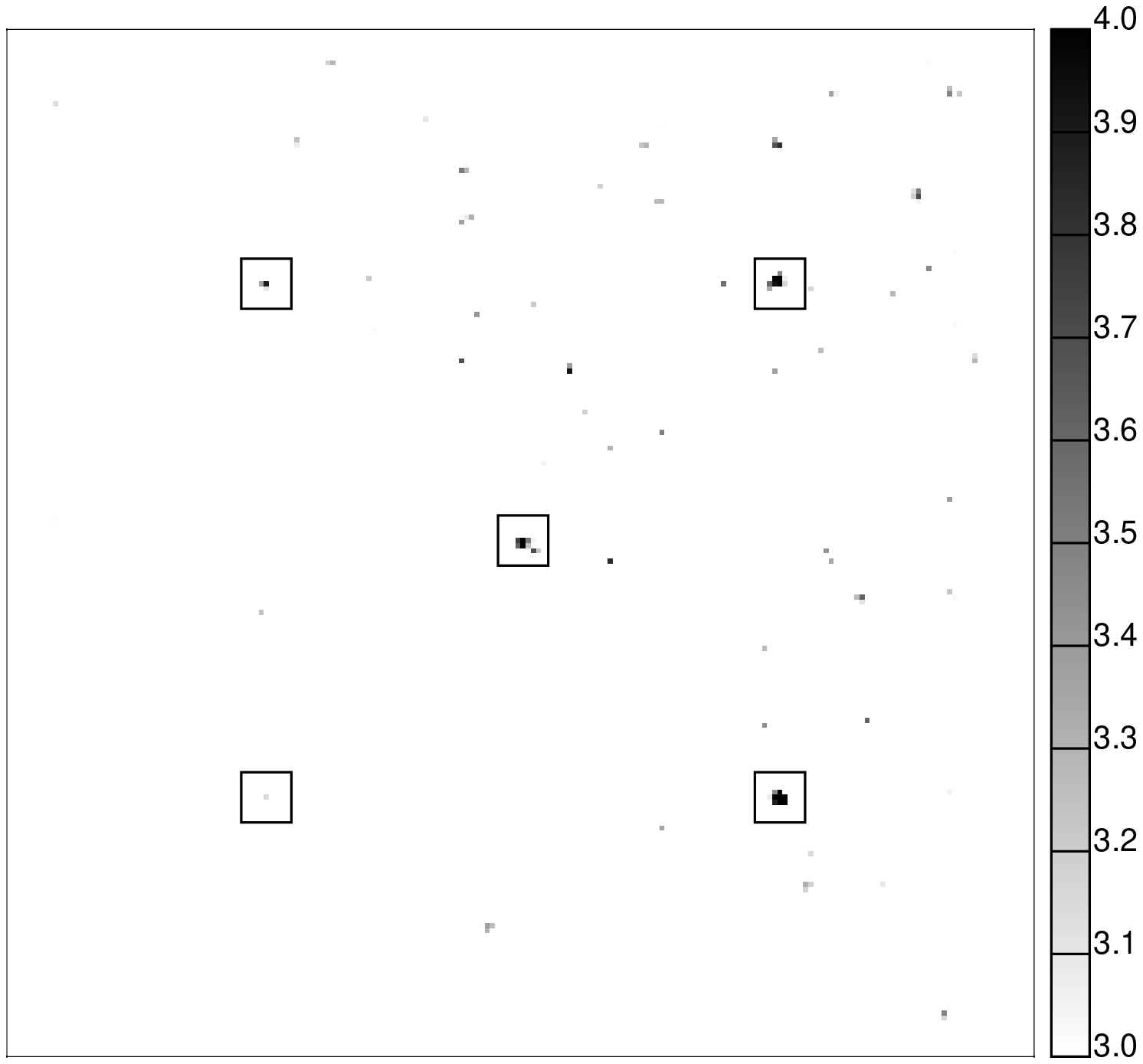}{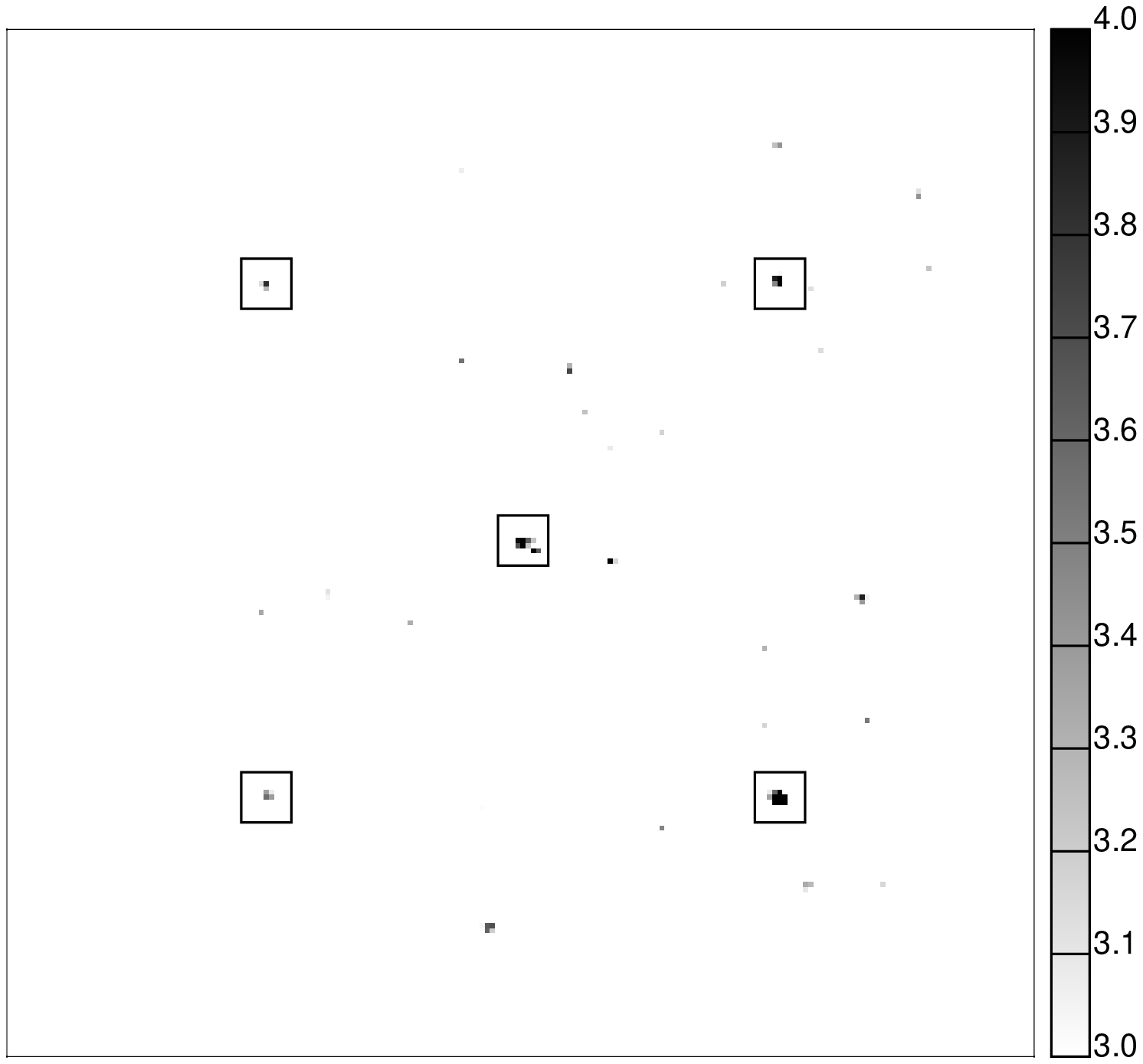}
\caption{
Significance map for the artificially generated cubic background map
shown in \fig{image}b, using {\bf (a)} a one-annulus detector with a
13x13 pixel background, and {\bf (b)} a two-annulus detector with a
29x29 inner annulus, and 41x41 outer annulus.  In both cases, the
source region was 3x3 and the annular hole was 5x5.  Point 1 (lower left)
is seen to be more significant in the two-annulus map than in the one-annulus
map.  Also, there is a preponderance of false alarms in the upper right
quadrant of the one-annulus map.}
\label{fig-cubic-amap}
\end{figure}

In \fig{cubic-amap}, we show significance maps for the cubic
background map, again using one- and two-annulus methods.  Point~1,
the lower left artificial point source that is at the bottom of the
trough, is barely detected by the one-annulus method
with a significance of just above 3.0.  
From the figure, it also appears
that the one-annulus detector exhibits a spatially uneven number of
false alarms, with most of the false alarms occuring in the upper
right quadrant where the background has a peak.  
By contrast, the two-annulus method produces
false detections that are spread more evenly over the map, and the
real source in the lower left quadrant is more readily identified.

To study this effect more systematically, we repeated this experiment
a thousand times, and the results are shown in
Table~\ref{table-cubic-sigmas}. 
The two-annulus detector was slightly better, on average, than the
one-annulus detector for points 2, 4, and 5; these are points on the
diagonal where the background curvature is small.  The two-annulus
detector was substantially better for point~1, which is in the the
trough.  The one-annulus detector reported a larger sigma
for point~3, the upper right point source, but that is also where the
one-annulus detector has its highest false alarm rate.  In fact, the
one-annulus detector reported an average significance (4.2 sigmas) for
this point that was larger than the nominal significance (4.0 sigmas)
that was designed into the experiment.

\subsection{Generating ``surrogate maps''}

It is sometimes possible to analytically characterize the false alarm
rate (or $p$-value) for simple point source detection algorithms when
applied to a single candidate point source; but it is generally a lot
more complicated, and often is outright intractable, to obtain false
alarm rates for entire maps.  In general, one discounts the
significance of a candidate point source by the number of independent
point source detection attempts (or ``trials'').  But when the source
regions and background annuli act as ``moving windows'' over the map,
the detection attempts are not independent, and it is nontrivial to
get the correct trials factor.
\citeN{Saha95} has discussed the spatial overlapping window problem,
and has computed (by numerical integration) the ``completeness''
versus ``spurious detection rate'' for a given magnitude limit and
signal to noise ratio.  \citeN{1994ApJ...423..714B} look at the
simpler case of overlapping time windows, and provide an empirical fit
to the trials factor (see Table 3 of that reference).  See
\citeN{Biller96} for a more extensive discussion of trial factors
in the context of combining statistical tests in an open-ended search.
If the
background is known precisely, and the sources are sufficiently rare,
the problem can be treated analytically~\cite{Politzer86}, but for our
unknown and spatially nonuniform backgrounds, we resort to Monte-Carlo
calculation.

If we have a good estimate of the background at each pixel, we can can
make an ensemble of ``surrogate maps'' by assigning each pixel a value
chosen from a Poisson distribution with mean equal to the background
at the pixel.  The surrogate maps, by definition, do not have point
sources, so any point source detections will be spurious.  Applying
the point source detection algorithm to a large number of such maps,
one can estimate of the false alarm rate, and in particular,
can find an appropriate threshold for the ``number of sigmas'' of significance
a single source must exhibit to be significant at a given level for the
entire map.

The most straightforward way to estimate the background at each pixel
is to use a one or two annulus estimator, as we have discussed in
earlier sections.  However, this generates a background map which is
not smoothly varying from pixel to pixel, but instead includes
fluctuation that arises from the Poisson statistics in the original
map.  Such surrogates generate more false alarms than are consistent
with the null hypothesis of a smoothly varying background.

Our approach for generating surrogate maps is to find a background
function $B(x,y)$ which is as smooth as possible while still being
consistent with the original map.  We can define ``smoothness'' in
terms of the two-annulus estimator.  A ``perfectly smooth'' map is
one that can be modeled exactly as a cubic polynomial; such a map 
will satisfy
\be	
	B_{ij} = \beta\binn+(1-\beta)\bout
\ee
where $\binn$ and $\bout$ are averages of $B_{ij}$ over the inner and outer
annuli, and $\beta$ is given by \eq{beta-twoannuli}.  Exact equality
is an unreasonable demand, because that implies a single cubic polynomial
fit over the entire background.  However, we can try to satisfy this
equality as closely as possible, while constraining the background to
be consistent with the data.  In particular, we want
\be
	\left\langle \frac{(N_{ij}-B_{ij})^2}{B_{ij}} \right\rangle 
	\approx 1.
	\label{eq-bg-eq-data}
\ee

To find a $B_{ij}$ which simultaneously satisfies both (approximate)
constraints, we take an iterative approach:
\begin{enumerate}
\item On the first iteration, we estimate $B_{ij}$ at each pixel using
      the ``average best'' linear combination of two annuli described
      in Section~\ref{sect-avgbest}.
\item On subsequent iterations, we refine this estimate of $B_{ij}$ 
      using the straight two-annulus method as defined in \eq{bcen-twoannuli}
      and \eq{beta-twoannuli}.  
\item For each of these subsequent iterations, however, we perturb
      our estimate by a small fraction $\epsilon$ so that it does not
      stray too far from the data.  That is,
\be
	B_{ij}^{(n+1)} = (1-\epsilon)\left(
	\beta\binn^{(n)}+(1-\beta)\bout^{(n)}\right)
	+ \epsilon N_{ij}.
\ee
      In our implementation, the $\epsilon$ is adjusted in an {\it ad hoc} 
manner until \eq{bg-eq-data} is satisfied; it is not difficult to imagine
ways to automate this selection.
\end{enumerate}

\begin{figure}
\plottwo{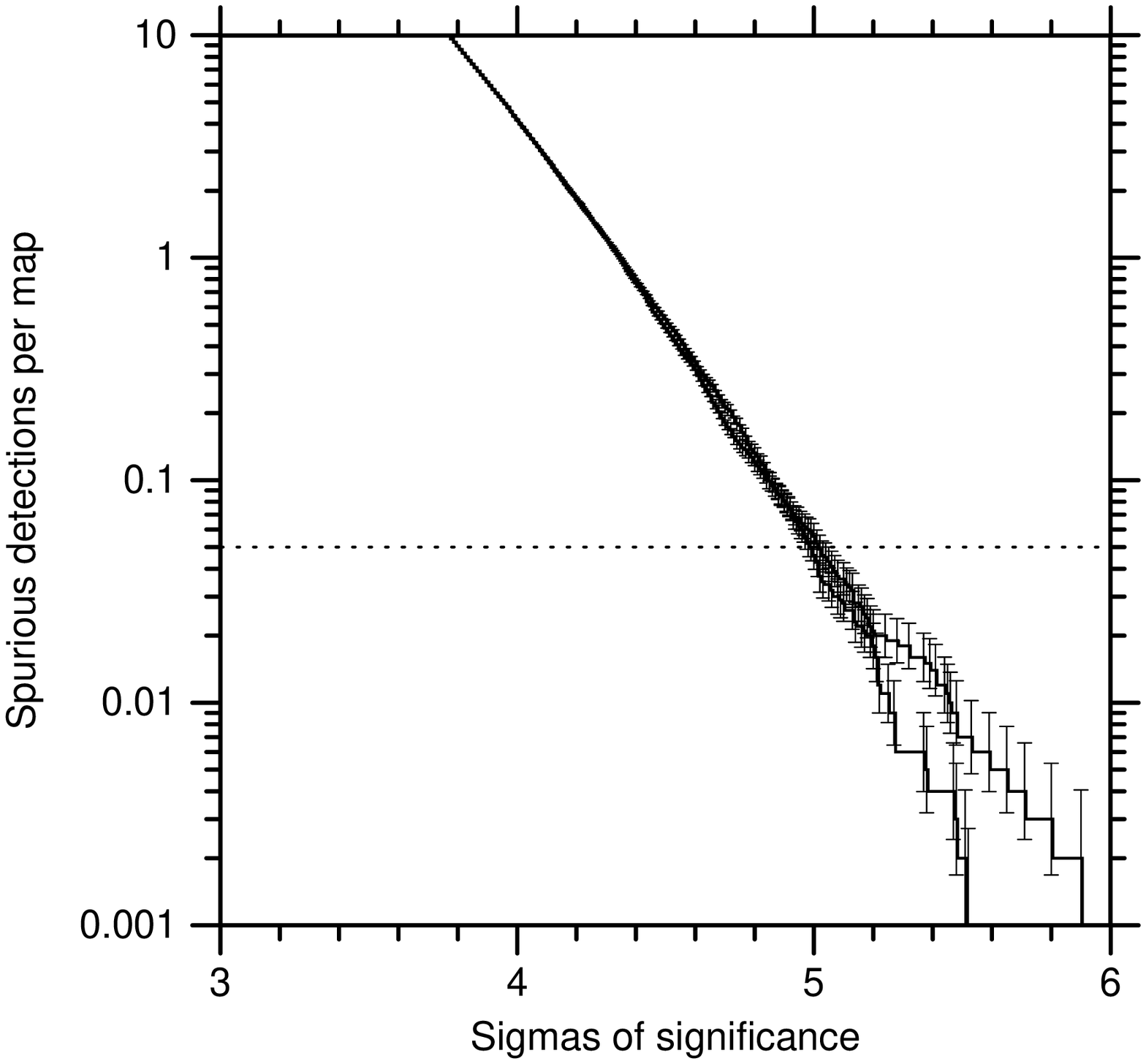}{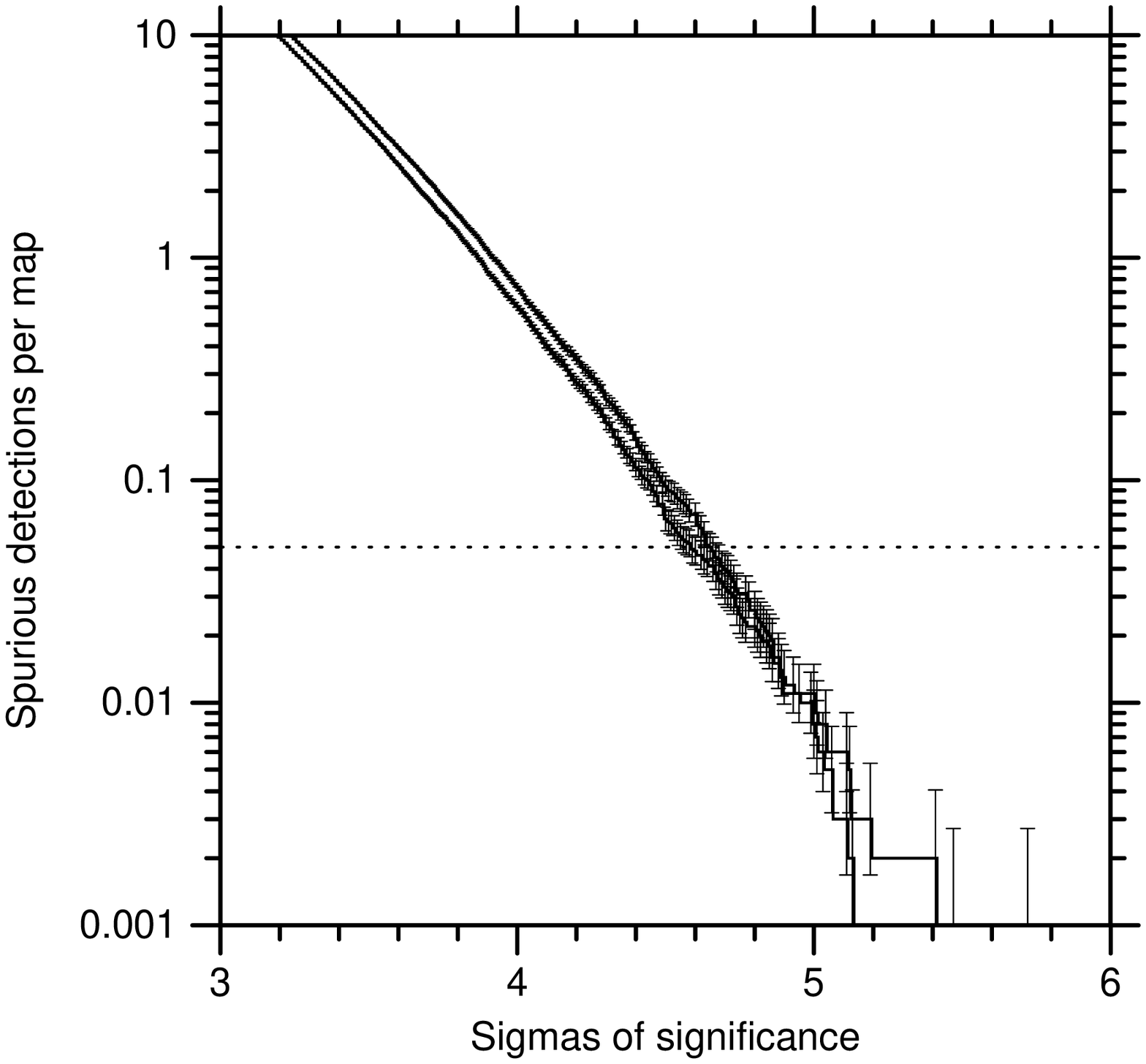}
\caption{Cumulative histograms generated from one thousand surrogate maps of 
both {\bf (a)} the ALEXIS map (\fig{image}a) 
and {\bf (b)} the artificial cubic background map (\fig{image}b),
using both one-annulus and two-annulus point-source
detection strategies (annulus sizes are described in the captions
to \fig{alexis-amap} and \fig{cubic-amap}). 
The horizontal axis is significance S/N, and the vertical axis 
is the average number of 
spurious sources per map which are significant at that level or greater.
The dotted line corresponds to the traditional p=0.05 level of significance.}
\label{fig-hist}
\end{figure}

This background map is computed just once; multiple surrogate maps are
generated by replacing each pixel of the background map with an
integer chosen randomly from a Poission distribution with mean equal
to the floating point background value.  Figure~\ref{fig-hist} shows the
results of one thousand surrogate maps; for each map, the point source
algorithm is applied, and the statistics of point source detections
(which by definition are all spurious) are maintained in a cumulative
histogram.  This histogram provides a calibration which tells how many
spurious detections are expected in the map as a function of ``sigmas
of significance'' for a single detection.  The dotted lines in
figure~\ref{fig-hist} indicate the required
number of sigmas for a single source detection in order to achieve the
traditional level of $p$=0.05 significance for the entire map.  For the
ALEXIS map, this value is 5.0 sigmas; for the artifical map, it is
about 4.6 sigmas.

\begin{figure}
\plottwo{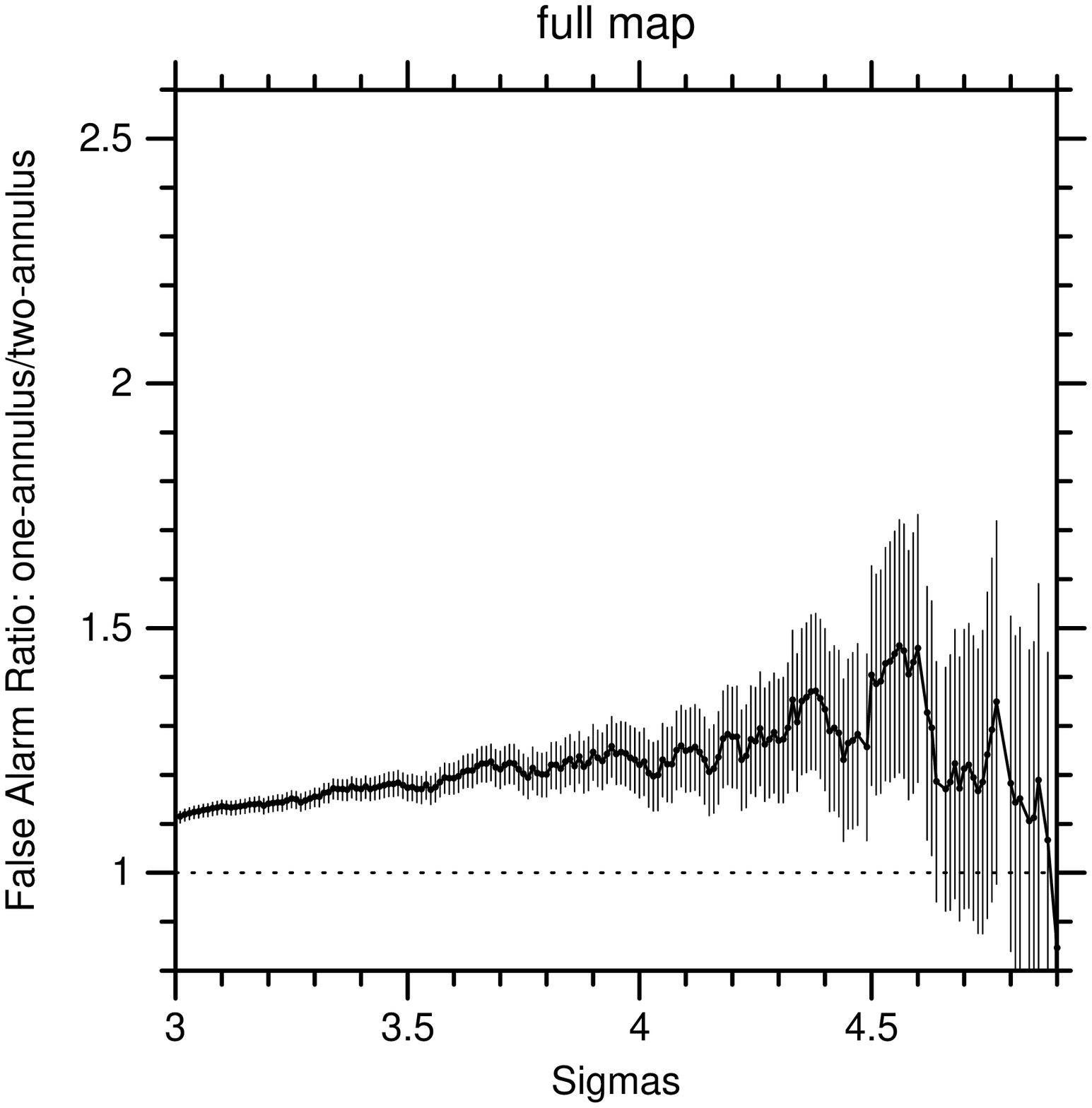}{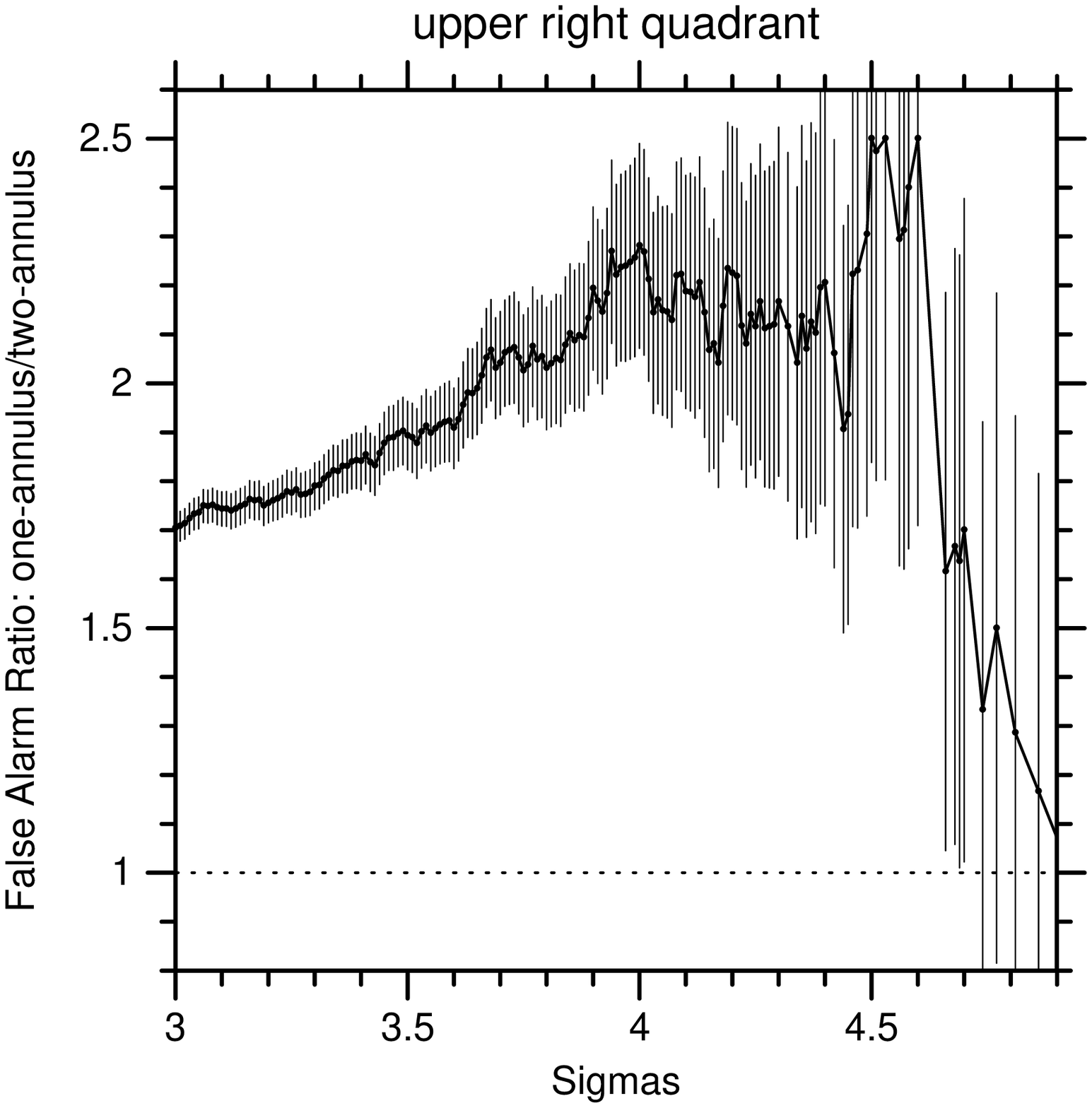}
\caption{{\bf (a)}
The ratio of false alarm rates for the one- and two- annulus detectors
is shown for the artificial cubic map (see \fig{hist}b).  
{\bf (b)} We restrict attention to the upper-right quadrant of the cubic map; 
in this
quadrant the background is peak-like, and we see that the false alarm rate
is roughly twice as large with the one-annulus detector compared to 
the two-annulus
detector.}
\label{fig-hist-ratio}
\end{figure}

From this calibration, we can compare the false alarm rates of the one
and two annulus detectors.  For the ALEXIS data, as seen in
\fig{hist}a, there is little difference between the two.  There is a
discernible difference for the artificial cubic background map; 
as shown in \fig{hist-ratio}a, 
the one-annulus method
produces approximately 20\% more false alarms.  
This is not in itself a problem for the one-annulus detector,
since the important thing is that the false alarm rate can be 
estimated; one simply adjusts the level of $S/N$ that is needed for
a confident detection.  
More worrying is that fact that the one-annulus method's false
alarms do not occur uniformly in the map; if we restrict our attention
to the upper right quadrant (where the background is generally peaked),
then \fig{hist-ratio}b shows 
that the false alarm rate for the one-annulus method 
is roughly double that of the two-annulus method.

\begin{figure}
\plottwo{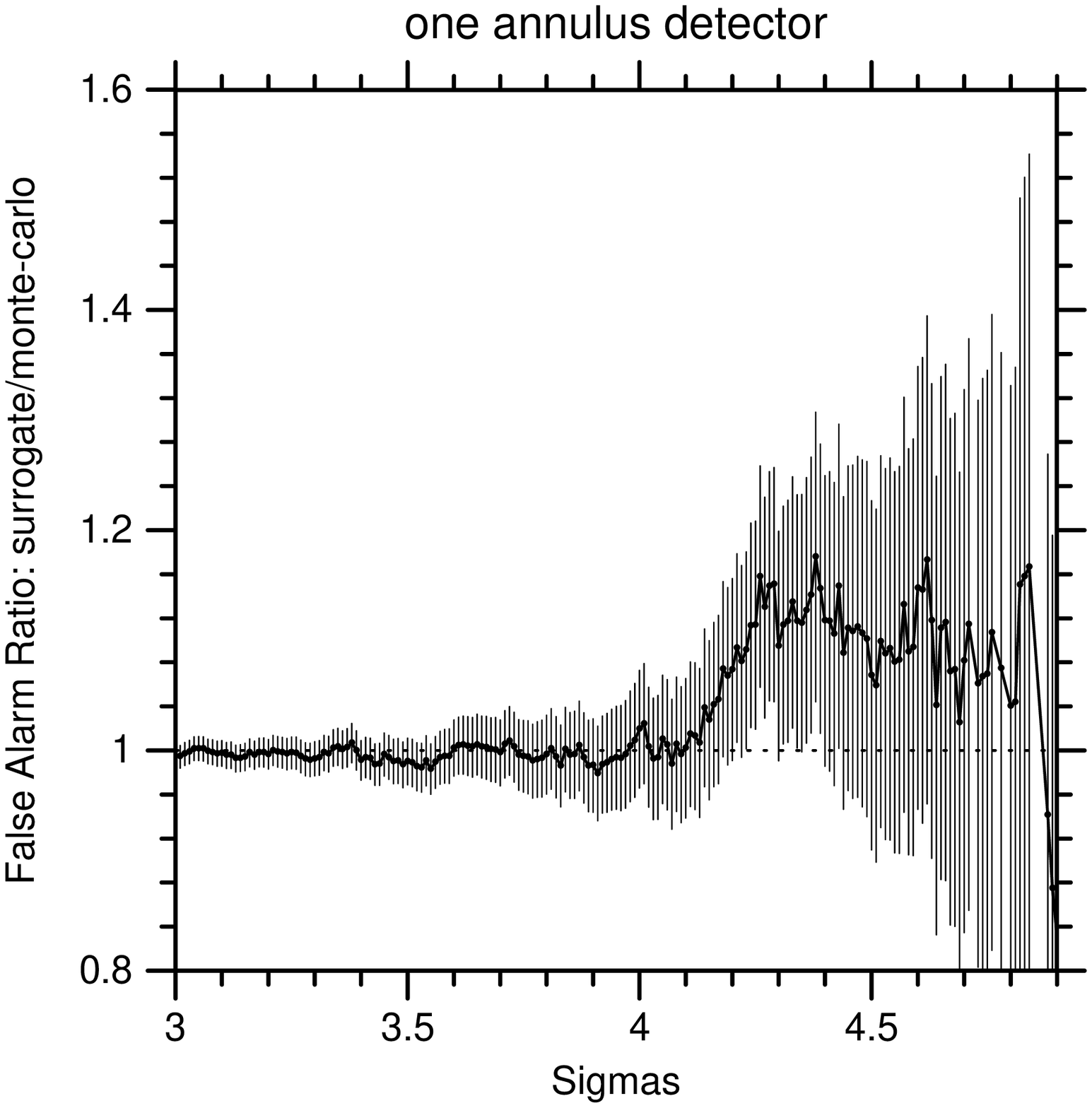}{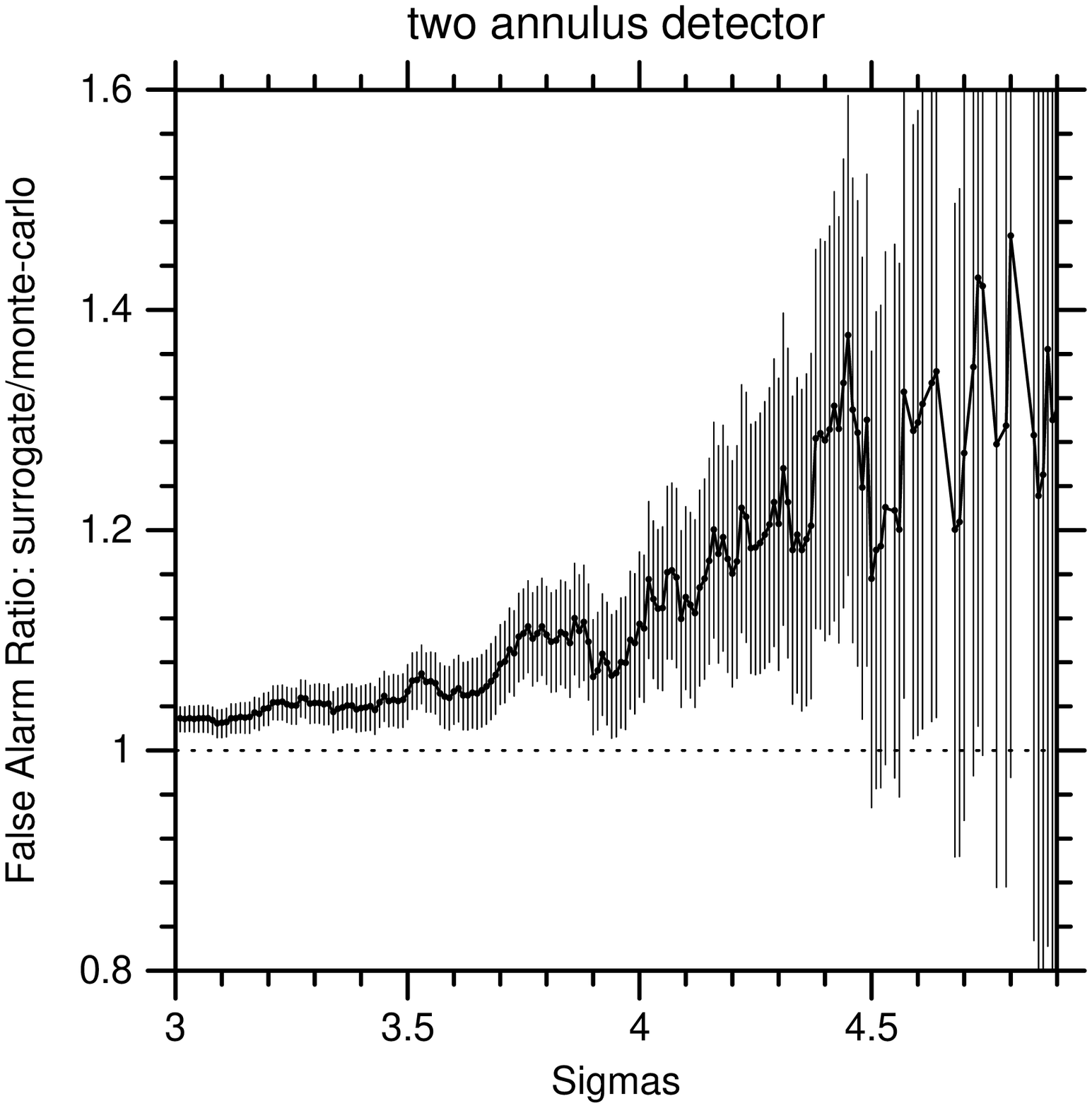}  
\caption{The false alarm rate estimated from 
the surrogate maps is divided by the ``true'' false alarm rate estimated from
a Monte-Carlo run with 1000 realizations of the artificial cubic-background map 
(with no artificially added point sources).  This ratio is plotted against
significance threshold for {\bf (a)} the one-annulus detector and {\bf (b)}
the two-annulus detector.} 
\label{fig-sur-test}
\end{figure}

Since we know the true background for the artificial map, we can use
this to asses how well the surrogate maps estimate the true false
alarm rates.  To estimate these true rates, we performed a Monte-Carlo
experiment with one thousand realizations of the artificial cubic
background map, but without any point sources added, and applied the
one-annulus and two-annulus detectors to these realizations.
\Fig{sur-test} shows the ratio of the false alarm rate estimated from
the surrogate maps to the true false alarm rate estimated from the
Monte-Carlo experiment.  This ratio is
very nearly one (to within estimation error) for the one-annulus
detector, and within a few percent of one for the two-annulus detector.
That the surrogate maps provide a slightly higher false alarm rate 
leads to slightly more conservative point source
detections in the real map.

\subsection{Characterizing background nonuniformity}
\label{sect-nonunif}
Although the multiple and continuum annuli estimators can account
for smooth (polynomial) background nonuniformity when the count density
is not too low, there are still times when a single-annulus estimator
is preferred.  As we've seen, a single-annulus estimator has lower
variance than a multiple-annulus estimator.  Under the assumption of
uniform background, one can furthermore obtain an exact test
of significance ({\it e.g.,} \citeNP{Lampton94}), 
even for arbitrarily low counts.
But even in this
single-annulus scenario, it is useful to check the assumption of
background uniformity.

One natural measure, which was originally used on the ALEXIS project,
is the Poisson dispersion index~\cite{Freund66}.  If $N_i$ is the count
in the $i$-th pixel in the background annulus, and there are a total
of $K$ pixels, then the statistic
\be
	X = \frac{\sum_{i=1}^K (N_i-\overline{n})^2}{\overline{n}}
	\label{eq-pdi}
\ee
where $\overline{n}=\sum_i N_i/K = N/K$ is the average counts per pixel,
will be approximately 
chi-square distributed with $K-1$ degrees of freedom.
If $X$ is large ($X\gg K-1$), that indicates a pixel-to-pixel 
nonuniformity that is larger than can be accounted for by Poisson 
variation.

This statistic is less than ideal from the point of view of point
source detection for several reasons.  Since there is no spatial
information coded into the statistic, it is only sensitive to
pixel-by-pixel variations.  This makes it relatively insensitive to a weak
nonuniformity that is spread over a large area; this is a problem that
is exacerbated when the pixels are small.  A related small-pixel
problem occurs when there are many more pixels than photons (when
$\overline{n}$ is much less than one).  In this regime, the
distribution of $X$ is no longer expected to be chi-square.  In fact,
in the limit as $\overline{n}\to 0$, each pixel will contain either zero
photons or one, and $X=(K^2-N^2)/N$ regardless of how those photons
are arranged.

On the other side of the coin, the Poisson dispersion index is
sensitive to all kinds of deviations from uniformity, including those,
such as a linear gradient, for which a single-annulus estimator is
quite adequate (is optimal in fact) for estimating the background in
the source area.  The statistic will (correctly) indicate that the
background is nonuniform, but a perfectly good background estimate
will be needlessly disqualified.

\begin{figure}
\plottwo{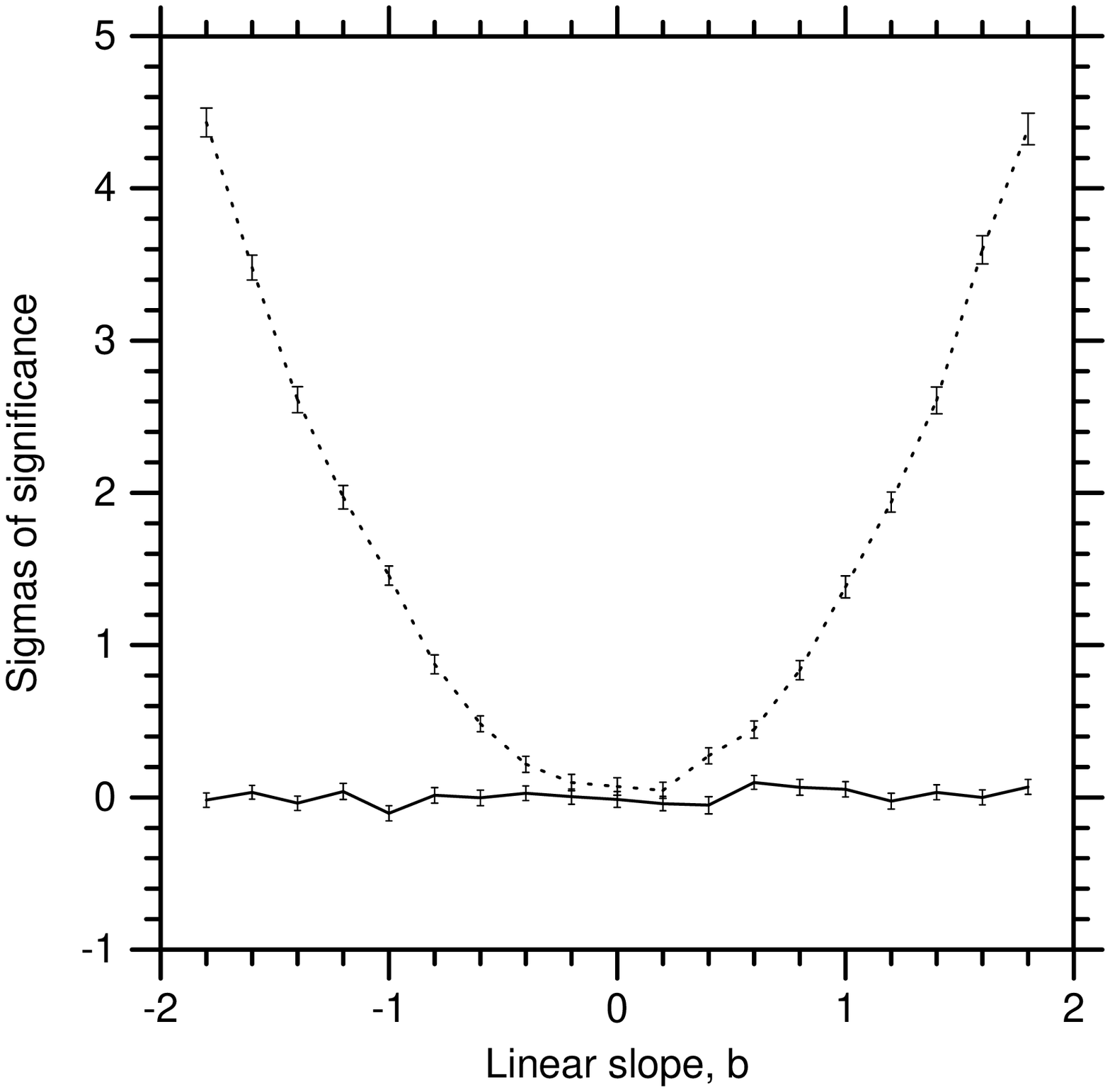}{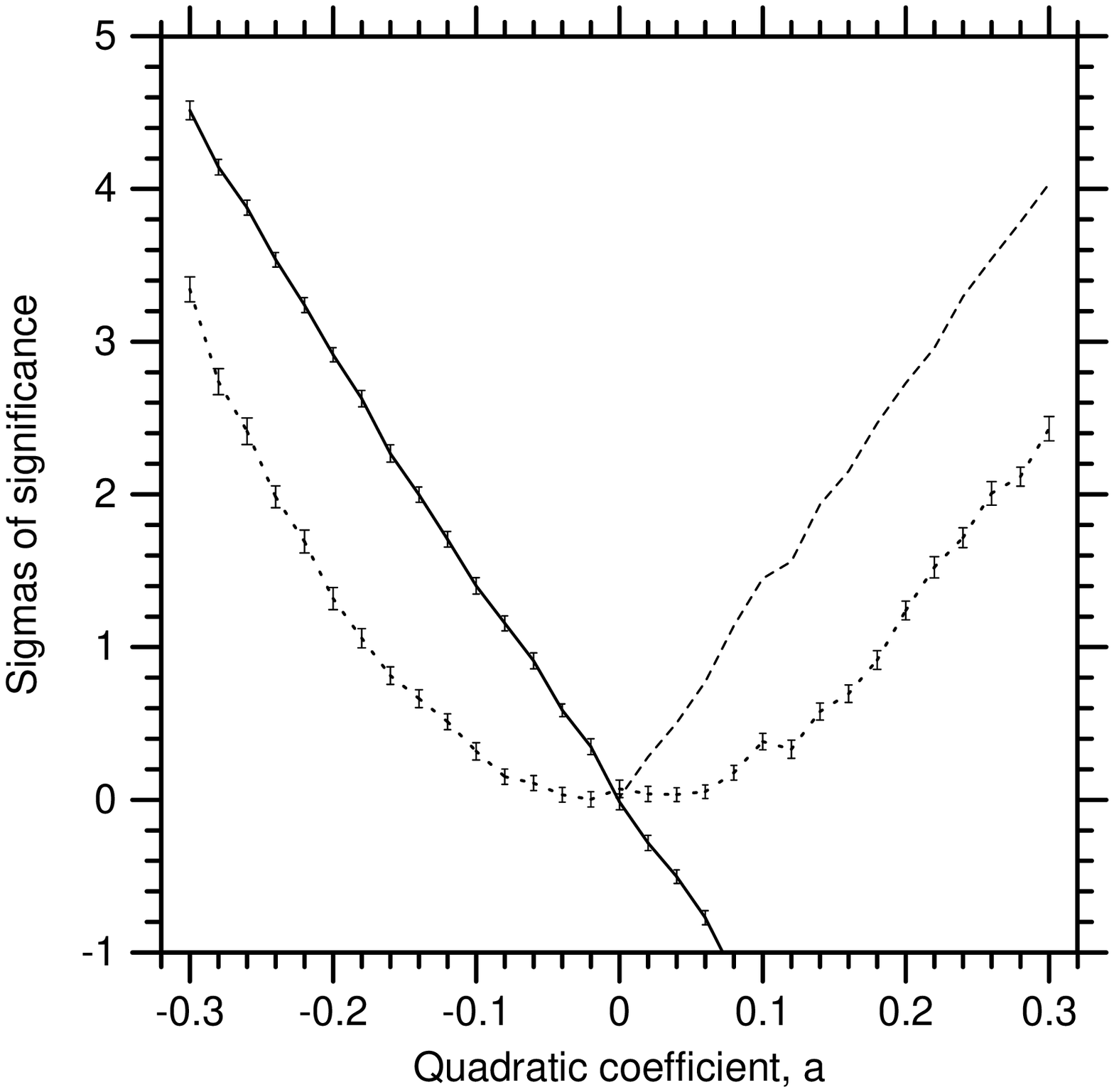}
\caption{Results of a Monte-Carlo experiment comparing the
sensitivity of the
Poisson Dispersion Index (PDI) (dotted line) in \protect{\eq{pdi}} with
an annulus-based statistic (solid line) in \protect{\eq{annulus-nonuniformity}},
over a range
of nonuniform backgrounds.  The background is generated by a quadratic equation:
$B_{ij} = c + bi + a(i^2+j^2)$, where $i,j$ are the pixel indices with values
$i=0$, $j=0$ corresponding to the center pixel of a 13x13 map.  For each trial,
a Poisson realization was constructed for the given background, and the
two-annulus statistic (using a 9x9 inner annulus) and the PDI was computed.  
Both statistics have mean zero and variance one when
the background is flat, and so both can be used as a number of 
sigmas of signficance.  Plotted is the mean (with error bars indicating the
standard deviation of the mean) value of each statisic for 400
trials.  In {\bf (a)}, the slope $b$ is varied while keeping the
quadratic coefficient $a=0$, and in {\bf (b)} the quadratic
coefficient $a$ is varied while keeping the slope $b=0$.  
In both cases $c=100$ counts.
We see from {\bf (a)} that the annulus-based
statistic is completely insensitive to simple linear gradients, whereas
the PDI is sensitive even to gradients of order one
percent per pixel.  (We do not mean to advocate the PDI as a sensitive
detector of gradients; the high sensitivity in this case is a function of 
the relatively large counts per pixel and the large number of pixels 
in the background annulus.)  By contrast, we see from {\bf (b)} that 
the annulus-based statistic is more
sensitive to small quadratic nonuniformities.  Note that a negative 
coefficient $a$ corresponds to a ``peak'' in nonuniformity, and 
these peaks in the background are just the artifact that are most likely
to lead to spurious source detections.  Positive values of $a$ (``valleys''
in the background) lead to
negative significance. If a two-tailed test of flatness is desired, then
the absolute value of the statistic should be used; this is shown as a
dashed line in this plot.}
\label{fig-bg-slope}
\end{figure}

Using concentric annuli, we will develop a statistic that is both more
sensitive to weak quadratic nonuniformities and at the same time
insensitive to such ``benign'' nonuniformities as linear gradients (or
any odd polynomial powers).%
\footnote{Actually, it is not a bad idea to directly 
employ the Poisson dispersion index in \eq{pdi}, but with annular 
regions in place of
individual pixels.  If the regions have different areas, then a slight
modification is needed, but such a statistic is by design insensitve
to linear gradients, and since annular regions are generally much larger
than individual pixels, one expects the chi-squared approximation to be
more accurate.}  We will define a statistic which is a simple 
linear function of the counts in the annuli:
\be
	T = \sum_s \eta_s N_s
	\label{eq-annulus-nonuniformity}
\ee
where the coefficients $\eta_s$
are chosen to satisfy the conditions
\bea
	\sum_s \eta_sA_s &=&0,\\
	\sum_s \eta_s^2A_s/\sum_sA_s &=& 1.
\eea
This statistic will have mean zero, and variance $\langle T^2 \rangle = N$,
where $N=\sum_s N_s$ is the total number of counts.  In particular, we
will be able to reject the null hypothesis of uniform background with
$T/\sqrt{N}$ ``sigmas'' of significance.  If there are two annuli, then
there is only one solution for $\eta_s$: $\eta_\inn = \sqrt{(1-\alpha)/\alpha}$,
and $\eta_\out = \sqrt{\alpha/(1-\alpha)}$.  For more than two annuli, we
note, without going into detail,
that it is possible to choose the coefficients $\eta_s$
so that the statistic is optimally sensitive to quadratic nonuniformities --
these are the nonuniformities that lead to the most serious mis-estimates
of background when the single annulus estimator is used (a continuum-limit 
statistic, in which the coefficients $\eta_s$ are replaced by
a function $\eta(r)$, is also straightforward to derive). \Fig{bg-slope}
compares the Poisson dispersion index and a two-annulus statistic, showing
that the two-annulus statistic is completely insensitive to to linear 
gradients, but is more sensitive to weak quadradic nonuniformities.

The last two columns of Table~\ref{table-cubic-sigmas} show how the annulus-based
background 
statistic is
able distinguish troughs (negative value at point~1), peaks (positive value
at point~3), and low curvature points (2, 4, and 5) in the background
nonuniformity.  
As with ordinary point source detection, there is a
trade in the choice of annulus size.  Choosing an outer annulus equal in
size to the annulus that is used for point source detection provides an
appropriately local measure of nonuniformity; choosing a larger annulus
size however provides a more sensitive measure.

In the special case that there are two background annuli, then the
distribution of ($N_\inn,N_\out$) is binomial and an exact $p$-value
can be computed in terms of the incomplete beta function.  
This was pointed out by 
\citeN{Lampton94} in the context of determining whether 
the counts in a source region are consistent with the counts in 
a single background annulus.
In
particular, a one-sided p-value is given by the probability of
seeing $N_\inn$ or more counts in the inner annulus.
\be
	p = \sum_{n=N_\inn}^{N}
	\left(\begin{array}{c} {N} \\ n \end{array}\right)\,
	\alpha^n(1-\alpha)^{N-n} \equiv I_\alpha(N_\inn,N_\out+1)
\ee
where $N=N_\inn+N_\out$, and $\alpha=A_\inn/(A_\inn+A_\out)$, and
$I_\alpha$ is the incomplete beta function.   The one-sided p-value will
be small when $N_\inn$ is unusually large; this occurs when the
background is ``peaked'' near the candidate point source location,
the very nonuniformity that is most likely to lead to a spurious point
source detection.

We remark that this statistic does not ``assume'' that the background
is smoothly varying, or well-modeled by a polynomial.  It is simply a
test of the null hypothesis that the background is uniform.  It will
detect many different kinds of deviations from uniformity, but is
designed to be particularly powerful against case that the background
has a smooth convexity (or ``peakiness'').

\section{Conclusion}

We have derived linear multiple-annulus estimators that
provide unbiased background estimation when the background is
spatially nonuniform but smoothly varying.  The approach works for
both circular and square background annuli, and can be conveniently
applied to photon event lists as well as to maps of the sky with
photons binned into square pixels.  One trades accuracy for precision
(lower bias for increased variance) in going to a multiple-annulus
estimator, and if the count rate is low or if the background is nearly
uniform, the trade becomes unfavorable.  Fortunately, one does not
have to depend on general trends or asymptotic results to inform this
trade; the empirical background error index, described in 
Section~\ref{sect-ebei}, provides a simple figure of merit that
can be computed directly from the data.  As well as comparing
distinct background estimation methods, it can be used to
define an optimal annulus size and/or an optimal linear combination of
the one and two annulus estimators.

We have seen that multiple-annulus background estimators can, in some
cases at least ({\it e.g.,} the cubic background map),
provide more robust point source detection.  But if the count density
in the map is not very high, the increased variance of the multiple-annulus
estimators may lead to a preference for simple one-annulus point source
detection.  But even in this regime, multiple-annulus methods can
play a supporting role.  We have
shown how a multiple-annulus smoothness condition can be used to
generate nonuniform ``surrogate maps'' which are can be used for assessing
false detection rates and significance thresholds.  We have also shown
how to use multiple annuli to characterize the nonuniformity of the
background; potential point sources that are detected at places on the
sky map with a high degree of nonuniformity will be more suspicious
than those that are detected in flatter regions.

At this writing, the ALEXIS satellite is still flying, and is still taking
data.  The ALEXIS daily point source detection effort is described by
\citeN{A_Dupre_ADASS_96}.  As the mission nears completion, work is in
progress on a final point source catalog.

\section*{Acknowledgements}

  We are grateful to all the dedicated ALEXIS operators who showed up
at all hours of the day and night to tend satellite downlinks, for
providing inspiration and data to the ten o'clock scholars who wrote
this paper.  We are particularly thankful to Diane Roussel-Dupr\'e for
numerous helpful suggestions on various aspects of this project and
of this manuscript.
 
  This work was conducted under the auspices of the United States Department
of Energy.


\clearpage

\newpage

\begin{deluxetable}{lrrrr}
\tablecolumns{5}
\tablewidth{0pt}
\tablecaption{Point source detections in the ALEXIS map with S/N$>$5\label{table-sigmas}}
\tablehead{\colhead{} & \multicolumn{2}{c}{Coordinates} & 
                        \multicolumn{2}{c}{Significance $S/N$} \\
\colhead{ID} & \colhead{RA} & \colhead{Dec} & 
\colhead{Two-annulus} & \colhead{One-annulus}}
\startdata
Sirius B        & 101.11 & $-$16.84 & 38.63 & 37.65 \nl
RE J0457$-$280  & 74.24 & $-$28.25 & 24.26 & 23.75 \nl
WD 0549$+$158    & 88.09 & 15.96 & 17.16 & 16.17 \nl
RE J0515$+$324   & 78.74 & 32.72 & 10.17 & 10.14 \nl
$\epsilon$ CMa\tablenotemark{a}  & 104.45 & $-$29.05 & 9.53 & 8.74 \nl
$\zeta$ Pup\tablenotemark{a}     & 120.68 & $-$39.96 & 9.17 & 9.34 \nl
RE J0723$-$274   & 110.73 & $-$27.80 & 8.52 & 8.55 \nl
$\zeta$ Ori\tablenotemark{a}     & 85.13 & $-$1.83 & 8.30 & 7.69 \nl
$\gamma$ Ori\tablenotemark{a}    & 81.20 & 6.42 & 7.25 & 7.05 \nl
RE 0623$-$374    & 95.85 & $-$37.65 & 6.20 & 6.55 \nl
$\beta$ CMa\tablenotemark{a}      & 95.51 & $-$18.14 & 6.15 & 5.88 \nl
$\epsilon$ Ori\tablenotemark{a} & 84.08 & $-$1.19 & 6.13 & 5.52 \nl
$\beta$ Ori\tablenotemark{a}     & 78.47 & $-$8.30 & 5.51 & 5.08 \nl
V471 Tau       & 57.39 & 17.39 & 5.11 & 5.12 \nl
\enddata
\tablenotetext{a}{These sources are bright O and B stars, and
are likely to have been detected because of known UV leaks 
in the detector filters, not because of radiation in the ALEXIS bandpass.
}
\end{deluxetable}

\begin{deluxetable}{ccccc}
\tablecolumns{8}
\tablewidth{0pt}
\tablecaption{Significance of artificial points in cubic 
background map\label{table-cubic-sigmas}}
\tablehead{
   \colhead{} & 
   \multicolumn{2}{c}{Significance S/N\tablenotemark{ac}} &
   \multicolumn{2}{c}{Background nonuniformity\tablenotemark{bc}} \\
\colhead{Point} & \colhead{Two-annulus} & \colhead{One-annulus} & 
        \colhead{Large Annuli} & 
	\colhead{Small Annuli}}
\startdata
1 & 3.55 & 3.11 & $-$49.67 & $-$1.50 \nl
2 & 3.82 & 3.77 &   \phs\phn0.01 & $-$0.06 \nl
3 & 3.98 & 4.20 &  \phs20.68 & \phs0.54  \nl
4 & 3.83 & 3.77 & \phs\phn0.02 & $-$0.00  \nl
5 & 3.84 & 3.70 &  \phs\phn0.03 & \phs0.08   \nl
\enddata
\tablenotetext{a}{If the background were known exactly, the nominal significance
for these artificial points would be $S/N=4.0$.}  
\tablenotetext{b}{This statistic is defined in Section~\ref{sect-nonunif}.
The large annuli are 29$\times$29 and 41$\times$41; the small annuli
are 9$\times$9 and 13$\times$13.}
\tablenotetext{c}{The statistical error for 
all the numbers reported in this table is $\pm 0.03$.}
\end{deluxetable}

\end{document}